\documentclass[aps,prb,twocolumn,showpacs,apsrev]{revtex4-1}
\usepackage[T1]{fontenc}
\usepackage{lmodern}
\usepackage[english]{babel}
\usepackage{amsmath}
\usepackage{amsfonts}
\usepackage{graphicx} 
\usepackage{comment}
\usepackage{caption}
\usepackage{subcaption}
\usepackage{xfrac}
\usepackage{color}
\usepackage{multirow}
\usepackage{hyperref}
\usepackage{braket}
\usepackage{pstricks}
\usepackage{pgfplots}
\pgfplotsset{compat=newest}
  \usetikzlibrary{plotmarks}
  \usetikzlibrary{arrows.meta}
  \usepgfplotslibrary{patchplots}
  \usepackage{grffile}
  \usepackage{amsmath}

\hypersetup{
	colorlinks = true,
	pdftitle = {},
	pdfauthor = {},
	pdfkeywords = {},
	linkcolor = blue,
	citecolor = blue,
	filecolor = black,
	urlcolor = magenta
}

\newcommand{\ee}{\mathrm{e}}
\newcommand{\img}{\mathrm{i}}

\graphicspath{{./graphics/}}

\begin{document}

\title{First-Principles-Based Strain and Temperature Dependent Ferroic Phase Diagram of SrMnO$_3$}
 
\author{Alexander Edstr\"om}
\author{Claude Ederer}
\affiliation{Materials Theory, ETH Z\"urich, Wolfgang-Pauli-Strasse 27, 8093 Z\"urich, Switzerland}


\begin{abstract}

Perovskite structure SrMnO$_3$ is a rare example of a multiferroic material where strain-tuning and/or cation substitution could lead to coinciding magnetic and ferroelectric ordering temperatures, which would then promise strong magnetoelectric coupling effects. Here, we establish the temperature and strain dependent ferroic phase diagram of SrMnO$_3$ using first-principles-based effective Hamiltonians. All parameters of these Hamiltonians are calculated using density functional theory, i.e., no fitting to experimental data is required. Temperature dependent properties are then obtained from Monte Carlo and molecular dynamics simulations. We observe a sequence of several magnetic transitions under increasing tensile strain, with a moderate variation of the corresponding critical temperatures. In contrast, the ferroelectric Curie temperature increases strongly after its onset around 2.5\,\% strain, and indeed crosses the magnetic transition temperature just above 3\,\% strain. Our results indicate pronounced magnetoelectric coupling, manifested in dramatic changes of the magnetic ordering temperatures and different magnetic ground states as function of the ferroelectric distortion. In addition, coexisting ferroelectric and ferromagnetic order is obtained for strains above 4\,\%. Our calculated phase diagram suggests the possibility to control the magnetic properties of SrMnO$_3$ through an applied electric field, significantly altering the magnetic transition temperatures, or even inducing transitions between different magnetic states.

\end{abstract}

\maketitle

\section{Introduction}

Multiferroic materials, with coexisting
magnetic and ferroelectric (FE) order, have attracted much attention over the last decades, due to promises of important technological applications, as well as fundamental scientific developments~\cite{Ramesh2007,Spaldin2017}.
In many cases, this coexistence is restricted to low temperatures, or one of the two order parameters emerges only as a by-product of the other, i.e., as an \emph{improper} or secondary order parameter. Examples are so-called type-II multiferroics~\cite{Khomskii:2009}, where the magnetic order breaks inversion symmetry, thus allowing for a small electric polarization, typically induced either through spin-orbit coupling or exchange-striction.
Even though in such cases an intimate coupling exists between the primary and the secondary order parameters, the smallness of the secondary order parameter makes potential applications of these materials challenging.

On the other hand, in so-called type-I multiferroics~\cite{Khomskii:2009}, both magnetic and ferroelectric order often coexist up to room temperature or above. In this case, both magnetic and ferroelectric order parameters are primary, and thus generally not small. Since the two types of ferroic order typically arise from different mechanisms, they are, in a first approximation, independent from each other. Nevertheless, the same coupling mechanisms as in type-II multiferroics (e.g., spin orbit coupling, exchange-striction), with the same characteristic coupling strengths, are also at play in these systems. Typically, this coupling strength is weak compared to the energy scales governing the primary order parameters, and thus, in most cases, no pronounced coupling effects can be observed.
However, close to the ferroic ordering temperatures, the relevant response functions either diverge or attain large values. In this case, even a moderate coupling can give rise to drastic effects. In particular, if the two ferroic ordering temperatures coincide, or are close to each other, a variety of highly interesting and potentially useful coupling phenomena are expected, such as, e.g., temperature-mediated magneto-electric coupling~\cite{PhysRevLett.114.177205} or multicaloric effects~\cite{planes_multical}.

Generally, the ferroelectric and magnetic ordering temperatures in type-I multiferroics do not coincide, due to the different underlying mechanisms. Nevertheless, the considerations outlined above suggest a promising route for designing multiferroic materials with strongly coupled magnetic and ferroelectric properties by tailoring their ferroic ordering temperatures. A highly interesting material in that context is SrMnO$_3$.

Using first principles calculations, Lee and Rabe~\cite{PhysRevLett.104.207204} predicted that this otherwise paraelectric (PE) and G-type antiferromagnetic (AFM) material becomes FE under biaxial tensile strain, which also leads to a series of magnetic transitions.
In particular, a ferromagnetic (FM) phase has been predicted to occur under high tensile strain, thus also presenting the highly appealing possibility to obtain the rare combination of both FM and FE order.
The appearance of strain-induced polar order in SrMnO$_3$ has been corroborated by recent experiments~\cite{Becher2015,acs.nanolett.5b04455,ADMI201601040,PhysRevB.97.235135}, with very recent work achieving high quality films with strain as large as 3.8\% \cite{PhysRevB.97.235135}.
Furthermore, several experiments also indicated the appearance of FM order in strained SrMnO$_3$~\cite{doi:10.1063/1.4960463,BAI201757}, although the increased formation of oxygen vacancies under tensile strain~\cite{marthinsen_2016} is believed to play a crucial role, here.

It has also been suggested from first principles calculations, that the closely related hypothetical compound BaMnO$_3$, isoelectronic to SrMnO$_3$ but with larger volume, would be ferroelectric~\cite{PhysRevB.79.205119}. This has stimulated both experimental~\cite{PhysRevLett.107.137601,ADMI201601040} and computational~\cite{PhysRevB.94.165106} work showing that a multiferroic phase can also be obtained through Ba-substitution in SrMnO$_3$.
Thus, the combination of strain engineering and Ba-substitution allows for a very rich ferroic phase diagram to be explored.
Since the FE critical temperature is usually highly sensitive to epitaxial strain (see, e.g., [\onlinecite{doi:10.1063/1.4930306}]), while the magnetic ordering temperature is expected to be less sensitive, this allows for potentially coinciding magnetic and FE ordering temperatures, resulting in pronounced magnetoelectric coupling phenomena.
Enhanced magnetoelectric coupling is also expected due to the fact that both magnetic and FE order are related to the same $B$-site Mn cation, which also manifests in strong spin-phonon coupling~\cite{PhysRevB.84.104440,PhysRevB.85.054417,PhysRevLett.107.137601}.

As described above, first principles-based calculations have been crucial in identifying SrMnO$_3$ as a promising multiferroic material.
Thereby, previous computational work has been mostly limited to zero temperature~\cite{PhysRevLett.104.207204,marthinsen_2016,PhysRevB.94.165106}. However, an understanding of the complete phase diagram, and of the strong coupling phenomena expected in the regime where magnetic and FE critical temperatures coincide, requires access to finite temperatures.
In the present work, we therefore combine first principles calculations using density functional theory (DFT) with Monte Carlo (MC) and molecular dynamics (MD) simulations, and explore the temperature and strain-dependent ferroic phase diagram of SrMnO$_3$, including both magnetic and ferroelectric degrees of freedom.

We start from zero temperature DFT calculations presented in Sec.~\ref{result.DFT}. We then construct a Heisenberg Hamiltonian with exchange parameters extracted from DFT, and perform MC simulations to assess finite temperature magnetism in Sec.~\ref{result.mag}. The calculated strain dependent Heisenberg exchange interactions also shed new light on the strain-induced sequence of magnetic transitions, while the effect of FE structural distortions on the exchange interactions provide a first step towards a unified description of the coupling between magnetism and ferroelectricity in this system.
MD simulations based on an effective Hamiltonian describing the polar soft-mode displacements~\cite{PhysRevB.49.5828,PhysRevLett.73.1861,PhysRevB.52.6301}, which is also constructed from DFT calculations, are then used to study the FE finite temperature properties in Sec.~\ref{sec.FE}. Finally, a discussion of the complete ferroic strain-temperature phase diagram is presented in Sec.~\ref{sec.PD}, and conclusions are summarized in Sec.~\ref{sec.concl}.

\section{Computational Methods}\label{sec.meth}

DFT calculations are performed with VASP~\cite{KRESSE199615,PhysRevB.49.14251,PhysRevB.47.558} and projector augmented wave (PAW) pseudopotentials~\cite{PhysRevB.50.17953,PhysRevB.59.1758}. The exchange-correlation functional is described with the PBEsol version of the generalised gradient approximation~\cite{PhysRevLett.100.136406} (GGA) and an additional Coloumb repulsion~\cite{PhysRevB.57.1505} of $U_\mathrm{eff}=3~\mathrm{eV}$ on the Mn d-electrons, as has been motivated in previous studies of SrMnO$_3$~\cite{Becher2015,marthinsen_2016}. A recent study concluded that the version of GGA employed here provides better agreement with available experimental data than other commonly used exchange-correlation functionals~\cite{PhysRevB.93.205110}. The plane wave energy cut-off is set to 680 eV. A grid of at least $7 \times 7 \times 7$ $\mathbf{k}$-points for the basic perovskite unit cell, or correspondingly for supercell calculations, is used with a Gaussian smearing method for the Brillouin zone integration. Born effective charges and dielectric constants were evaluated using density functional perturbation theory\cite{PhysRevB.73.045112}. 

Using the Heisenberg exchange interactions evaluated according to the procedure described in Sec.~\ref{mag_par}, classical Metropolis~\cite{doi:10.1063/1.1699114} MC simulations are performed using the Uppsala Atomistic Spin Dynamics code~\cite{Skubic2008}, for $20 \times 20 \times 20 $ unit cells (8000 magnetic Mn atoms) and periodic boundary conditions, taking into account first and second nearest neighbor exchange interactions. A cooling process is simulated with 5 K increments, 20000 MC sweeps for thermalisation, and 150000 measurement sweeps at each temperature. A calculation for a $30\times30\times30$ system is performed to confirm that finite size effects are negligible. Neither the type of ordering nor the critical temperature, within the accuracy of a few Kelvins considered here, is affected by the increase in system size. 

To go beyond the zero temperature DFT results and access also finite temperature FE properties and phase transitions, an effective Hamiltonian approach~\cite{PhysRevB.49.5828,PhysRevLett.73.1861,PhysRevB.52.6301}, describing low energy structural distortions in terms of strain and soft mode variables, is used. This Hamiltonian is studied using an MD solver implemented in the FERAM code~\cite{PhysRevB.78.104104}. The parameters required as input for the effective Hamiltonian are obtained with DFT total energy calculations or density functional perturbation theory, largely following the scheme described in~[\onlinecite{PhysRevB.82.134106}], and discussed further in Sec~\ref{sec.FE}. The MD simulations are performed on a $32 \times 32 \times 32$ supercell. This is done in the canonical ensemble, using a Nos\'{e}-Poincar\'{e} thermostat~\cite{BOND1999114}. A time step of $2~\mathrm{ps}$ is used and 50000 time steps are performed in the thermalisation phase and 200000 in the measurement phase. Strained bulk is simulated by fixing the in-plane homogenous strain variables as has been done in previous studies of strain effects on ferroelectricity~\cite{doi:10.1063/1.4879840,doi:10.1063/1.4930306}. This allows one to study the effect of, e.g., a film clamped to a substrate, while focusing on bulk effects rather than finite size surface effects. Most results are presented for calculations where the system is initialised in a homogenously polarised FE phase at low temperature and then gradually heated, although calculations are also been performed for a cooling procedure starting from a random configuration above the critical temperature. 

\section{Zero Temperature Strain-Dependent Calculations}\label{result.DFT}

Our DFT calculations for SrMnO$_3$ as a cubic perovskite G-type antiferromagnet result in a lattice parameter of $a_0=3.79~\AA$, which agrees with the experimental value of $3.805~\AA$~\cite{PhysRevB.64.134412}. Neutron diffraction experiments have indicated that SrMnO$_3$ is G-type AFM with a Mn magnetic moment of $2.6 \pm 0.2 \mu_\mathrm{B}$ at liquid nitrogen temperature, in accord with our calculated value of $2.8\mu_\mathrm{B}$. Previous computational studies~\cite{PhysRevLett.104.207204,marthinsen_2016,Becher2015} have also indicated small tilts of the oxygen octahedra, which will be neglected here, as has been motivated before~\cite{Becher2015} and is, furthermore, consistent with experimental studies that report a cubic structure.

\begin{figure}[hbt!]
	\centering
	\includegraphics[width=0.49\textwidth]{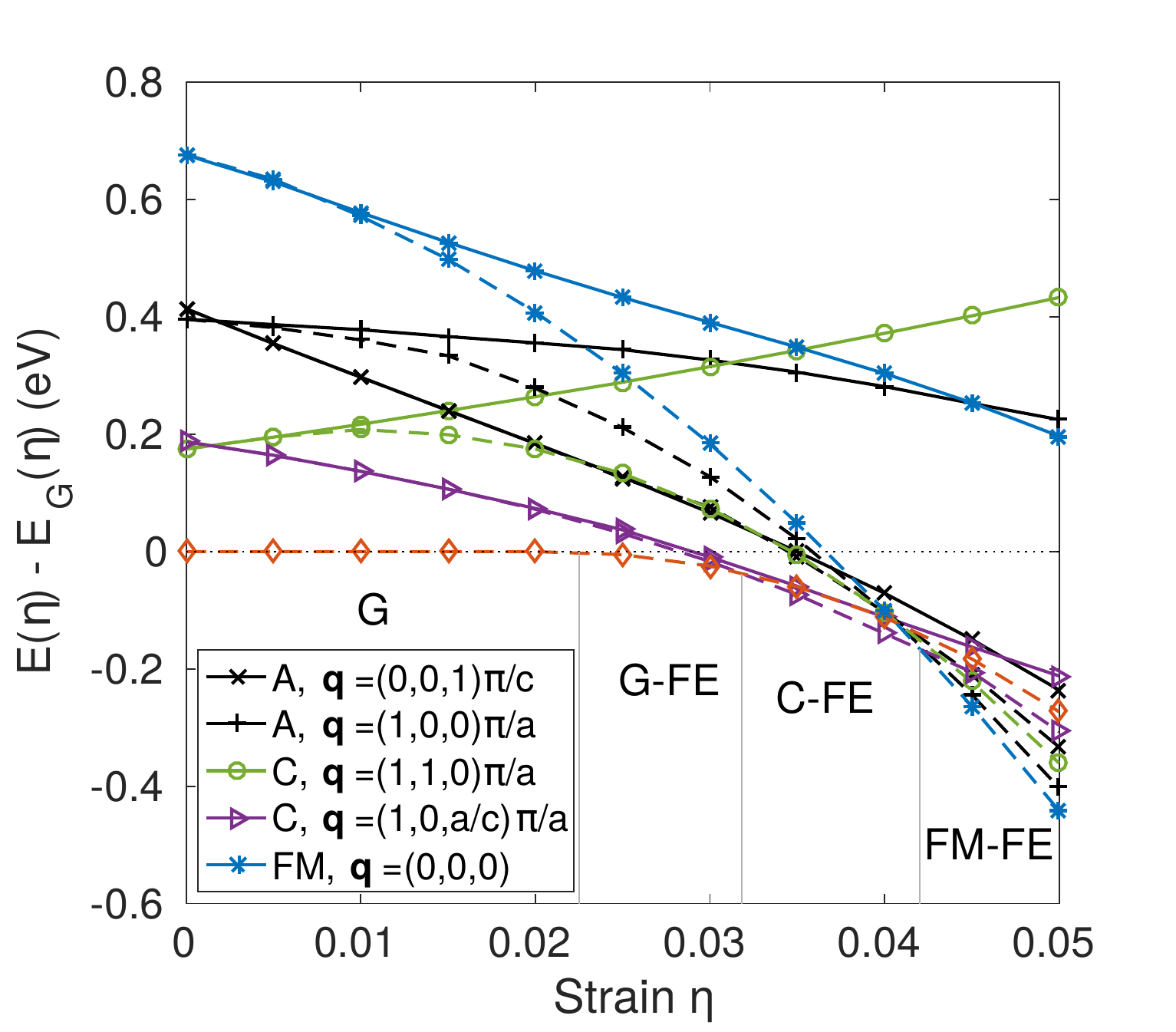}
	\caption{Energy as a function of strain for different magnetic orders, initialized with (dashed lines) or without (solid lines) polar structural distortions. The red dashed curve with diamond shapes shows the energy of the G-type AFM structure relaxed from the initialized polar structure. The energies are presented relative to that of a centrosymmetric G-type AFM structure, corresponding to zero energy, at each strain. }
	\label{fig.Eofa}
\end{figure}

Following [\onlinecite{PhysRevLett.104.207204}], we first address the emergence of ferroelectricity under tensile epitaxial strain at zero temperature within our computational setup.
Biaxial tensile strain ($\eta = a/a_0 - 1$) is considered, where the strain indicates the increase in the fixed in-plane lattice parameter relative to $a_0$, while the out-of-plane lattice parameter $c$ is allowed to relax.
Fig.~\ref{fig.Eofa} shows the total energy as function of strain, computed with DFT, for different magnetic ordering (A, C and G-type AFM~\cite{Wollan/Koehler:1955}, as well as FM).
For each magnetic order, we consider two cases: one (indicated by solid lines in Fig.~\ref{fig.Eofa}) where we constrain the structure to remain centrosymmetric (with space group $P4/mmm$), and one (dashed lines in Fig.~\ref{fig.Eofa}) where we initialize small polar displacements along the (110)-direction (space group $Amm2$). 
If the solid and dashed lines coincide, this indicates that the polar structure has relaxed back to the centrosymmetric one, and that FE order is not favored with the given magnetic structure and strain.
The energies are given relative to that of the centrosymmetric structure with G-type AFM at each strain.

In the cubic structure, A-type and C-type AFM correspond to $\mathbf{q}$-vectors $(1,0,0)\frac{\pi}{a}$ (equivalent to $(0,1,0)\frac{\pi}{a}$ and $(0,0,1)\frac{\pi}{a}$) and $(0,1,1)\frac{\pi}{a}$ (equivalent to $(1,0,1)\frac{\pi}{a}$ and $(1,1,0)\frac{\pi}{a}$), respectively. However, in the strained structure, $(1,0,0)\frac{\pi}{a}$, is no longer equivalent to $(0,0,1)\frac{\pi}{c}$, and similarly for C-type, resulting in two inequivalent A-type and C-type magnetic orders. Each of these are included in Fig.~\ref{fig.Eofa}, with the types of ordering labeled by the corresponding $\mathbf{q}$-vectors. Note also that this symmetry breaking of the magnetic order results in a slightly different value for the $c$-lattice parameter for the different A-type orderings ($3.79~\AA$ and $3.82~\AA$ for $\mathbf{q}$-vectors $(0,0,1)\frac{\pi}{c}$ and $(1,0,0)\frac{\pi}{a}$, respectively), and thus also slightly different total energies at zero strain in Fig.~\ref{fig.Eofa} (we remind that the in-plane lattice parameters are kept fixed at a strain relative to the relaxed cubic lattice parameter with G-type AFM, while the out-of-plane lattice parameter is allowed to relax).

In the intervall 0-2\% strain, the centrosymmetric structure with G-type AFM is lowest in energy, while from 2.5\% strain and up, the non-centrosymmetric structures become lower in energy. Furthermore, C-type AFM, corresponding to $\mathbf{q}=(1,0,a/c)\frac{\pi}{a}$, becomes lowest in energy in the range 3.5-4\% strain. Here it should be mentioned that all relaxed polar structures show atomic displacements along the (110)-direction, except in the case of C-type AFM order with $\mathbf{q}=(1,0,a/c)\frac{\pi}{a}$, for which they are along the (100)-direction, even though these calculations were initialized with small symmetry-breaking displacements along (110). This can be understood in terms of the symmetry breaking between the (100) and (010)-directions by this specific type of magnetic order, and it is essentially a manifestation of the magnetically induced phonon anisotropy, which has been discussed in this material by others~\cite{PhysRevB.84.104440}. The result that the atomic displacements occur in a direction along which the Mn-atoms are antiferromagnetically coupled, rather than that along which Mn-atoms are ferromagnetically coupled, is consistent with the previous studies of magnetically induced phonon anisotropy in SrMnO$_3$~\cite{PhysRevB.84.104440}. For large strain, in the range 4.5-5\%, FM ordering becomes lowest in energy. It is interesting to note that the FM state is only favored at large strain if the structure is allowed to turn FE, whereas in the centrosymmetric structure FM ordering remains high in energy for all strains considered. This fact will be discussed further in terms of the Heisenberg exchange interactions in the next section.

The results in Fig.~\ref{fig.Eofa} can be compared to the results presented in [\onlinecite{PhysRevLett.104.207204}], which, however, were obtained using a different GGA functional ([\onlinecite{PhysRevLett.104.207204}] uses the PBE~\cite{PhysRevLett.77.3865} version instead of PBEsol~\cite{PhysRevLett.100.136406} used here). 
The main trend of going from G-PE, to G-FE, and then via C-FE to FM-FE is very similar. However, according to the calculations in [\onlinecite{PhysRevLett.104.207204}], FE order emerges at a lower strain of about 1\%, and a FE A-type AFM phase appears in a narrow strain range between the C-FE and FM-FE regions. These differences can mostly be ascribed to the larger equilibrium volume obtained using PBE, effectively renormalizing the corresponding strain values. In addition, a somewhat different $U_\mathrm{eff}=1.7~\mathrm{eV}$ ($U_\mathrm{eff}=U-J$ with $U=2.7~\mathrm{eV}$ and $J=1.0~\mathrm{eV}$) used in [\onlinecite{PhysRevLett.104.207204}], compared to the $U_\mathrm{eff}=3.0~\mathrm{eV}$ used here, can also lead to some differences. 
We note that a recent comparative study discussed the choice of functional and its effect on the calculated properties of Sr$_{1-x}$Ba$_x$MnO$_3$ and found that PBEsol yielded results in better agreement with experiments than PBE~\cite{PhysRevB.93.205110}.  

\section{Magnetism}\label{result.mag}

In order to assess temperature-dependent magnetic properties, we now discuss the magnetic ordering in terms of exchange interactions and the Heisenberg Hamiltonian. First, the evaluation of the Heisenberg exchange interactions from DFT is discussed in Sec.~\ref{mag_par}. Then, the results of MC simulations, using these exchange interaction parameters as input, are discussed in Sec.~\ref{mag_MCres}. 

\subsection{Calculation of Heisenberg Exchange Parameters $J_{ij}$}\label{mag_par}

For studying finite temperature magnetism, including the strain-induced transitions between different magnetic orders, the magnetic energy is mapped on a classical Heisenberg Hamiltonian,
\begin{equation}\label{eq.Heis}
H = - \sum_{i<j} J_{ij} \hat{m}_i \cdot \hat{m}_j \quad , 
\end{equation}
with exchange interactions $J_{ij}$ and unit vectors $\hat{m}_i$ describing the directions of the magnetic moments. Note that Eq.~\ref{eq.Heis} uses a sign convention such that positive $J_{ij}$ favors positive spin alignment (ferromagnetism). The Heisenberg exchange parameters are obtained from DFT calculations of the total energy for different collinear spin configurations, according to 
\begin{equation}
  \label{eq.Jij}
J_{ij} = \frac{E_{\uparrow \uparrow} + E_{\downarrow \downarrow} - E_{\uparrow \downarrow} - E_{\downarrow \uparrow}}{4n}, 
\end{equation}
with the arrows indicating the directions (up or down) of a given pair of magnetic moments ($i$ and $j$) in the corresponding spin configuration, and $n$ is equal to the number of equivalent bonds between atoms $i$ and $j$ that appear due to the finite size of the used supercell~\cite{PhysRevB.84.224429,PhysRevB.91.165122}. Calculations are performed for a $2 \times 2 \times 2 $ supercell, allowing up to third nearest neighbor interactions to be taken into account. 

When computing the magnetic exchange interactions according to Eq.~\ref{eq.Jij}, a magnetic reference state for the spins other than the considered pair must be chosen. For an ideal Heisenberg system, the exchange interactions are independent of the chosen reference state, while in real materials some differences are expected~\cite{PhysRevB.91.165122}. All exchange interactions presented in this paper are computed with respect to a G-type AFM reference state. For comparison, calculations of the exchange interactions as functions of strain were also performed for A and C-type AFM and FM reference states (in each case using identical lattice parameters). While this leads to small shifts in the exchange interactions as functions of strain, reflecting that SrMnO$_3$ is not an ideal Heisenberg system, the qualitative trends remain unchanged~\cite{suppl}. 

In the cubic structure the calculated first, second, and third nearest neighbor exchange interactions are $J_1=-13.10~\mathrm{eV}$, $J_2=-0.90~\mathrm{eV}$, and $J_3=0.22~\mathrm{eV}$. This can be compared to previously reported calculated values of $J_1=-13.95~\mathrm{eV}$, $J_2=-0.72~\mathrm{eV}$, and $J_3=0.01~\mathrm{eV}$~\cite{PhysRevB.84.104440} (note that in [\onlinecite{PhysRevB.84.104440}] a different definition of the Heisenberg Hamiltonian was used and therefore different values are given there). The agreement between these values and the ones reported here is rather good, and the small disagreement can be attributed to different values for $U_\text{eff}$, different exchange-correlation functionals, and slightly different computational schemes to evaluate the exchange interactions.

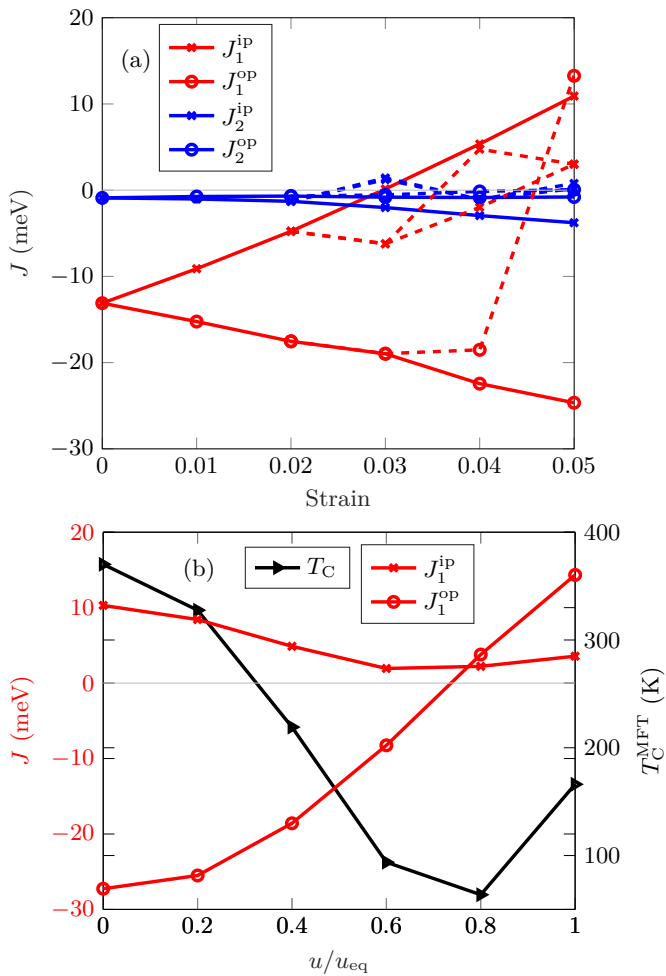
\begin{figure}[hbt!]
	\centering
	\hspace{-0.66cm}
%
%
\begin{tikzpicture}

\begin{axis}[%
width=0.35\textwidth,
height=0.32\textwidth,
at={(0\textwidth,0\textwidth)},
scale only axis,
xmin=0,
xmax=0.05001,
xlabel style={font=\color{white!15!black}},
xlabel={Strain},
xtick = {0, 0.01, 0.02, 0.03, 0.04, 0.05},
x tick label style={/pgf/number format/fixed},
scaled ticks=false,
ymin=-30,
ymax=20,
ylabel style={font=\color{white!15!black}},
ylabel={$J$ (meV)},
axis background/.style={fill=white},
legend style={at={(0.12,0.643)}, anchor=south west, legend cell align=left, align=left, draw=white!15!black}
]
\addplot [color=red, line width=1.3pt, mark=x, mark options={solid, red}]
  table[row sep=crcr]{%
0	-13.1024585000006\\
0.01	-9.11394900001028\\
0.02	-4.77185325000562\\
0.03	0.134391499997832\\
0.04	5.35742325000399\\
0.05	10.9154647499992\\
};
\addlegendentry{$J_{1}^\mathrm{ip}$}

\addplot [color=red, line width=1.3pt, mark=o, mark options={solid, red}]
  table[row sep=crcr]{%
0	-13.1033700000103\\
0.01	-15.2374177499937\\
0.02	-17.5280902499892\\
0.03	-18.9754202500012\\
0.04	-22.438219499989\\
0.05	-24.6551441249991\\
};
\addlegendentry{$J_{1}^\mathrm{op}$}

\addplot [color=blue, line width=1.3pt, mark=x, mark options={solid, blue}]
  table[row sep=crcr]{%
0	-0.895188937494851\\
0.01	-1.02195499999524\\
0.02	-1.29162374999581\\
0.03	-2.01594324999732\\
0.04	-2.94825356250072\\
0.05	-3.77497125000303\\
};
\addlegendentry{$J_{2}^\mathrm{ip}$}

\addplot [color=blue, line width=1.3pt, mark=o, mark options={solid, blue}]
  table[row sep=crcr]{%
0	-0.894845499995967\\
0.01	-0.735608624992778\\
0.02	-0.686619062499005\\
0.03	-0.809403750004378\\
0.04	-0.851782062508022\\
0.05	-0.786765562502012\\
};
\addlegendentry{$J_2^\mathrm{op}$}

\addplot [color=red, dashed, line width=1.3pt, mark=x, mark options={solid, red}, forget plot]
  table[row sep=crcr]{%
0.02	-4.77185325000562\\
0.03	-6.21940399999943\\
0.04	-1.98836512500833\\
0.05	3.01312774998763\\
};
\addplot [color=red, dashed, line width=1.3pt, mark=x, mark options={solid, red}, forget plot]
  table[row sep=crcr]{%
0.02	-4.77185325000562\\
0.03	-6.21940399999943\\
0.04	4.75623387499269\\
0.05	3.01312774998763\\
};
\addplot [color=red, dashed, line width=1.3pt, mark=o, mark options={solid, red}, forget plot]
  table[row sep=crcr]{%
0.02	-17.5280902499892\\
0.03	-18.9567613750015\\
0.04	-18.5140878750119\\
0.05	13.2664776249882\\
};
\addplot [color=blue, dashed, line width=1.3pt, mark=x, mark options={solid, blue}, forget plot]
  table[row sep=crcr]{%
0.02	-1.29162374999581\\
0.03	1.41925531249854\\
0.04	-1.07950431250003\\
0.05	0.185531000006733\\
};
\addplot [color=blue, dashed, line width=1.3pt, mark=x, mark options={solid, blue}, forget plot]
  table[row sep=crcr]{%
0.02	-1.29162374999581\\
0.03	1.22507218750201\\
0.04	-1.07950431250003\\
0.05	0.778701000008652\\
};
\addplot [color=blue, dashed, line width=1.3pt, mark=o, mark options={solid, blue}, forget plot]
  table[row sep=crcr]{%
0.02	-0.686619062499005\\
0.03	-0.546230437500128\\
0.04	-0.141121749997808\\
0.05	0.0622461250046058\\
};

\node[right, align=left]
at (axis cs:0.001,15.0) {(a)};

\addplot[mark=none,color=gray!50] coordinates {(0,0) (1,0)};

\end{axis}
\end{tikzpicture}%
  \\
%
%
\begin{tikzpicture}
\pgfplotsset{set layers}

\begin{axis}[%
width=0.35\textwidth,
height=0.28\textwidth,
at={(0\textwidth,0\textwidth)},
scale only axis,
xmin=0,
xmax=1,
xlabel style={font=\color{white!15!black}},
xlabel={$u/u_\mathrm{eq}$},
separate axis lines,
every outer y axis line/.append style={black},
every y tick label/.append style={font=\color{black}},
every y tick/.append style={black},
ymin=50,
ymax=400,
ylabel={$T_\mathrm{C}^\mathrm{MFT}$ (K)},
axis background/.style={fill=white},
yticklabel pos=right,
legend style={at={(0.3,0.85)}, anchor=south west, legend cell align=left, align=left, draw=white!15!black}
]

\addplot [color=black, line width=1.3pt, mark=triangle, mark options={solid, rotate=270, black}]
  table[row sep=crcr]{%
0	370.205225742734\\
0.2	327.629106932668\\
0.4	219.177599081105\\
0.6	93.8122639932138\\
0.8	63.6376103296485\\
1	166.102780363086\\
};
\addlegendentry{$T_\mathrm{C}$}

\end{axis}

\begin{axis}[%
width=0.35\textwidth,
height=0.28\textwidth,
at={(0\textwidth,0\textwidth)},
scale only axis,
xmin=0,
xmax=1,
xlabel style={font=\color{white!15!black}},
separate axis lines,
every outer y axis line/.append style={black},
every y tick label/.append style={font=\color{red}},
every y tick/.append style={black},
ymin=-30,
ymax=20,
ylabel={\textcolor{red}{$J$ (meV)}},
axis background/.style={fill=white},
yticklabel pos=left,
legend style={at={(0.546,0.757)}, anchor=south west, legend cell align=left, align=left, draw=white!15!black}
]
\addplot [color=red, line width=1.3pt, mark=x, mark options={solid, red}]
  table[row sep=crcr]{%
0	10.2977978750118\\
0.2	8.42378862500937\\
0.4	4.8745033749924\\
0.6	1.93801662499027\\
0.8	2.22082212498975\\
1	3.56725637500688\\
};
\addlegendentry{$J_1^\mathrm{ip}$}

\addplot [color=red, line width=1.3pt, mark=o, mark options={solid, red}]
  table[row sep=crcr]{%
0	-27.2571687500047\\
0.2	-25.5017938750086\\
0.4	-18.5819116249988\\
0.6	-8.25014937500868\\
0.8	3.7841596249919\\
1	14.3359492499968\\
};
\addlegendentry{$J_1^\mathrm{op}$}

\node[right, align=left]
at (axis cs:0.15,15.0) {(b)};

\addplot[mark=none,color=gray!50] coordinates {(0,0) (1,0)};

\end{axis}
\end{tikzpicture}%
	\caption{In-plane (ip) and out-of-plane (op) exchange interactions as functions of (a) strain and (b) ferroelectric displacement $u$. The latter is calculated at 5\,\% strain and for lattice parameters fixed to the corresponding centrosymmetric structure. In (a) solid lines denote the exchange interactions calculated for the lowest energy centrosymmetric structure at a given strain, while the dashed lines are computed for the lowest energy polar structures, where these are lower in energy than the non-polar ones. In (b), the mean field estimate of the critical temperature is also shown for the given exchange interactions.}
        \label{fig.Jofa}
\end{figure}

Fig.~\ref{fig.Jofa}(a) shows the calculated Heisenberg exchange interactions for first ($J_1$) and second ($J_2$) nearest neighbors as functions of strain. Third nearest neighbor interactions were also computed, but are several times smaller than the second nearest neighbor interactions, and, furthermore, they are relatively insensitive to strain. Hence, they will be neglected in the following. In the cubic structure there is only one first and one second nearest neighbor interaction, with six and twelve-fold  coordination, respectively. In the strained structure, these differ depending on whether they are in-plane (ip) or out-of-plane (op). The solid lines indicate exchange interactions computed for the tetragonal centrosymmetric structure lowest in energy at each strain, i.e. using the structural parameters corresponding to the lowest solid line in Fig.~\ref{fig.Eofa} at each strain (but keeping the magnetic reference state as G-type AFM). The dashed lines show the exchange interactions computed for the lowest energy FE structures at each strain. The reason for the splitting of the FE $J_1^{\mathrm{ip}}$ at 4\% strain is that the polarization along the (100)-direction (which occurs due to the C-type magnetic ordering) causes inequivalent bonds parallel or perpendicular to this direction. In this case, the positive $J_1^\mathrm{ip}$ corresponds to the (010)-direction, for which there is a change in the relevant Mn-O-Mn bond angle, while the negative $J_1^\mathrm{ip}$ corresponds to the (100)-direction, for which the bond angle remains $180^\circ$. A similar albeit smaller splitting occurs also for $J_2^{\mathrm{ip}}$ at 3 and 5\% strain. 

The first nearest neighbor exchange interactions are dominant at most strains, and one can understand the changes in magnetic ordering with strain mainly in terms of these. In the cubic structure,  $J_1^{\mathrm{ip}}=J_1^{\mathrm{op}}$ are strongly negative, resulting in G-type AFM, with all nearest neighbor spins aligned antiparallel to each other. Considering first the centrosymmetric structures, there is a nearly linear decrease (increase in magnitude) in $J_1^{\mathrm{op}}$. This increase in the strength of the out-of-plane interaction can be expected, as the out-of-plane bond distance shrinks under tensile strain. $J_1^{\mathrm{ip}}$, on the other hand, decreases in magnitude and eventually changes sign. At large strain one thus expects the Mn spins in the centrosymmetric structures to be aligned parallel along the in-plane direction and and antiparallel along the out-of-plane direction, corresponding to A-type AFM, in agreement with the solid lines in Fig.~\ref{fig.Eofa}.
The region around 3\% strain, where C-type AFM order emerges in Fig.~\ref{fig.Eofa} (with $\mathbf{q}=(1,0,a/c)\frac{\pi}{a}$), corresponds to the region where $J_1^{\mathrm{ip}}$ changes sign and thus has a magnitude comparable to, or even smaller than, the AFM second nearest neighbor coupling, $J_{2}^{\mathrm{ip}}<0$. This is indeed consistent with the $\mathbf{q}=(1,0,a/c)\frac{\pi}{a}$ C-type order with antiparallel alignment of all out-of-plane nearest and in-plane second nearest neighbor spins.

SrMnO$_3$ is an insulating transition metal oxide. As such, the magnetic coupling is expected to be mediated by a superexchange mechanism~\cite{Anderson196399}, with antiferromagnetic exchange for $180^\circ$ Mn-O-Mn bond angles. It is thus surprising to see the change in sign of $J_1^\mathrm{ip}$ with strain in Fig.~\ref{fig.Jofa}, as the bond angle remains $180^\circ$. 
However, the idea of superexchange is based on an idealised model that is not expected to hold perfectly for real materials. Furthermore, it should be noted that the magnetic interactions here also show some degree of non-Heisenberg behaviour (see supplementary material~\cite{suppl}). Recent work has also suggested that the Heisenberg or non-Heisenberg behaviour of exchange interactions can be heavily influenced by symmetries of the orbitals involved~\cite{PhysRevLett.116.217202}. Additionally, previous studies~\cite{PhysRevLett.116.217202,PhysRevLett.84.3169} have discussed how different competing exchange mechanisms can have different distance dependencies, which can potentially lead to a sign change such as that seen in $J_1^{\mathrm{ip}}$ in Fig.~\ref{fig.Jofa}.  Future investigation into the precise mechanism of the sign change in $J_1^\mathrm{ip}$ is therefore of great interest. 

As can be seen from the dashed lines in Fig.~\ref{fig.Jofa}(a), the FE structural distortion alters the exchange interactions in addition to the strain. In particular, at 5\% strain there is a drastic change in $J_1^{\mathrm{op}}$, including a change in sign. Thus, taking into account the FE distortion results in both of the dominating nearest neighbor interactions to be positive, favoring FM over AFM order.
In the FE structure, the atoms are shifted from their high symmetry positions, but also the out-of-plane lattice parameter changes slightly. Fig.~\ref{fig.Jofa}(b) shows how $J_1$ varies as the atomic positions are gradually changed from their high symmetry positions ($u=0$) to the equilibrium FE positions (characterized by a displacement amplitude $u$, defined such that $u=0$ corresponds to the centrosymmetric structure and $u=u_\mathrm{eq}$ corresponds to the relaxed non-centrosymmetric structure), while keeping the lattice parameters fixed to the equilibrium values for the FE structure with 5\% strain ($c=3.70~\AA$). At $u=0$, the $J_1$ are very similar to those seen for the centrosymmetric structure at 5\% strain in Fig.~\ref{fig.Jofa}(a). This implies that the effect of the change in lattice parameter is very small, as might be expected since the out-of-plane lattice parameter $c$ changes less than a hundredth of an \AA ngstr\"om. 
Instead, it appears that the change in the Mn-O-Mn bond angle causes the considerable change in $J_1^{\mathrm{op}}$. As $u$ varies from zero to $u_\mathrm{eq}$, this bond angle changes by $16^\circ$, from $180^\circ$ to $164^\circ$. Considering a superexchange mechanism, one would expect that going from a $180^\circ$ towards a $90^\circ$ bond angle favors ferromagnetism over antiferromagnetism. However, although the change in $J_1^{\mathrm{op}}$ follows the expected trend for a superexchange mechanism, considering the relatively small variation in bond angle, the change in the exchange interaction is surprisingly drastic here. Again, it appears that further investigations into the details of the exchange mechanisms for this material are of great interest. 

Fig.~\ref{fig.Jofa}(b) also shows the mean field estimate~\cite{MFTnote,Anderson196399,PhysRevB.70.024427} of the critical temperature calculated from the first nearest neighbor interactions as function of $u$. The critical temperature initially decreases with $u$, since the magnitude of both $J_1^{\mathrm{ip}}$ and $J_1^{\mathrm{op}}$ decrease in magnitude. It then reaches a minimum near where $J_1^{\mathrm{ip}}$ changes sign, after which it increases slightly again. Nevertheless, both $J_1$ are decreased in magnitude at $u=u_\mathrm{eq}$ compared to $u=0$, resulting, in total, in a decrease in the critical temperature. The transition from A-AFM to FM order occurs in the range $0.6 \leq u/u_\mathrm{eq}\leq 0.8$ according to the mean field model with first nearest neighbour interactions, i.e. in the range where $J_1^{\mathrm{ip}}$ changes sign.

\subsection{Monte Carlo Simulations of the Heisenberg Hamiltonian}\label{mag_MCres}

\begin{figure}[hbt!]
	\centering
	\includegraphics[width=0.5\textwidth,trim={0mm 0  0mm 0},clip]{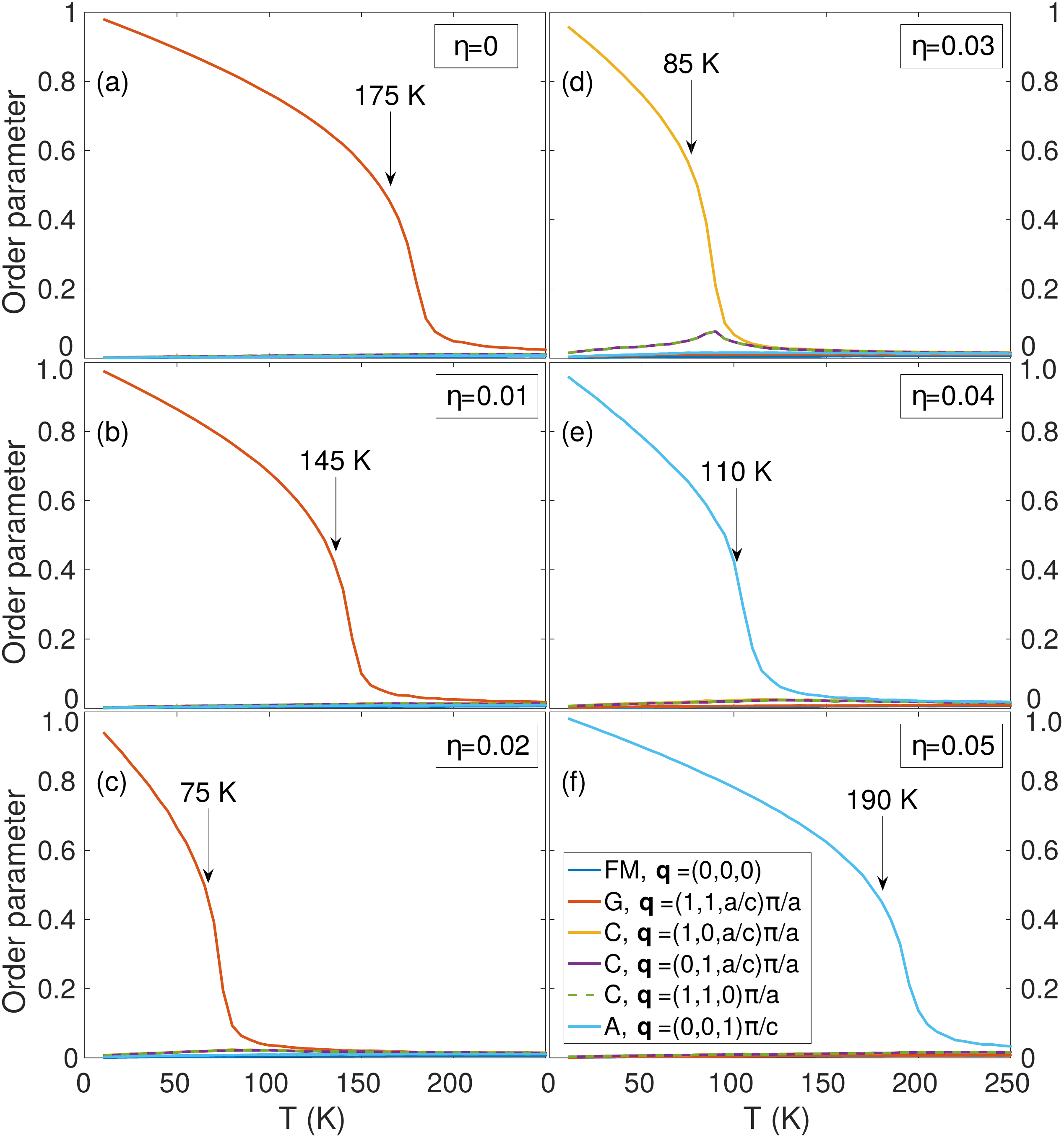} 
	\caption{Different magnetic order parameters as functions of temperature for tensile strains of 0, 1\%, 2\%, 3\%, 4\% and 5\% in (a)-(f), respectively. The critical temperatures, determined from the corresponding peak in the specific heat, are indicated in the plots. Note that each order parameter is included in every subfigure, but in each case only one is non-zero and clearly visible.}
	\label{fig.magMC_strain}
\end{figure}

Fig.~\ref{fig.magMC_strain}(a)-(f) contains results of MC simulations for the Heisenberg Hamiltonian, Eq.~\ref{eq.Heis}, using the first and second nearest neighbor exchange interactions computed from the centrosymmetric tetragonal structures (solid lines in Fig.~\ref{fig.Jofa}) at different strains.
Order parameters for different magnetic orders are plotted as function of temperature. Thereby, the (collinear) magnetic order parameter $M_\mathbf{q}$, corresponding to reciprocal space vector $\mathbf{q}$, is defined as: 
\begin{equation}
M_\mathbf{q} = \frac{1}{N}\sum_i \ee^{\img \mathbf{q}\cdot \mathbf{R}_i} m_i, 
\end{equation}
where $\mathbf{R}_i$ is the position of the $i$th magnetic moment out of $N$, and $m_i$ is its projection on the spin quantization axis. Typically $\ee^{\img \mathbf{q}\cdot \mathbf{R}_i}=\pm 1$ for $\mathbf{q}$ on the Brillouin zone boundary. 

According to the results in Fig.~\ref{fig.magMC_strain}, centrosymmetric tetragonal SrMnO$_3$ is a G-type AFM from 0 to 2\% strain, C-type AFM at 3\% strain and A-type AFM at larger strain. This sequence of transitions agrees with the zero temperature DFT results presented as solid lines in Fig.~\ref{fig.Eofa}. However, the critical strain for the transition from C to A-type AFM differs, as C-type AFM is lowest in energy at 4-4.5\% strain according to Fig.~\ref{fig.Eofa} while A-type AFM is favored already at 4\% strain in Fig.~\ref{fig.magMC_strain}e). This is likely due to some degree of non-Heisenberg behaviour in the system.
The MC simulations also confirm that no other magnetic structures, more complicated than the FM and AFM structures discussed so far, appear in any of the cases considered. This is also consistent with the adiabatic magnon spectra having minima only at the Brillouin zone boundary (see supplementary material~\cite{suppl}). In the cubic case it is known that a Heisenberg Hamiltonian with second nearest neighbour interactions does not show any non-collinear magnetic phases~\cite{10.1063/1.367729}.

\begin{figure}[hbt!]
	\centering
	\includegraphics[width=0.44\textwidth]{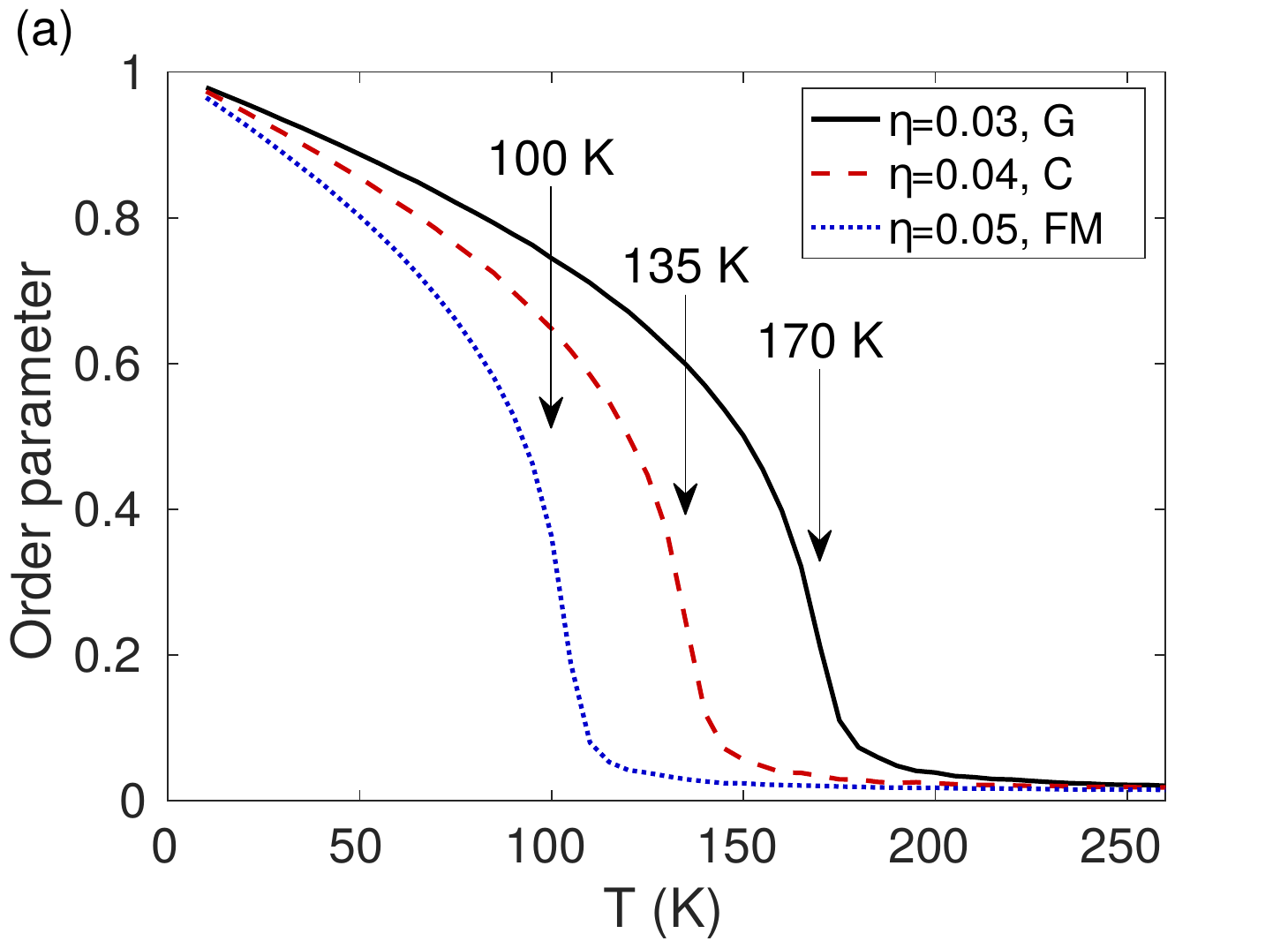}
	\includegraphics[width=0.44\textwidth]{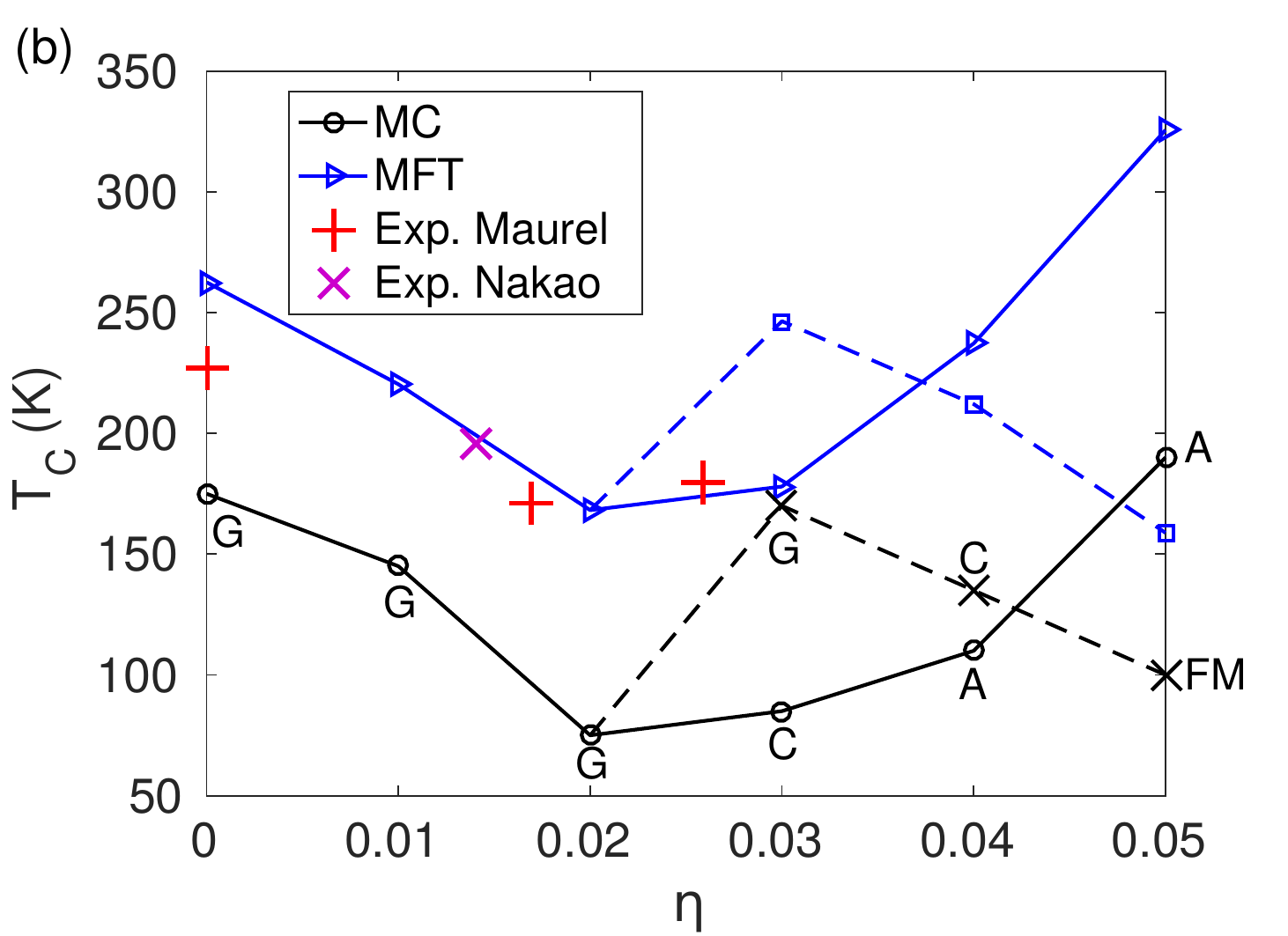} 	
	\caption{(a) Magnetic order parameters as functions of temperature for tensile strains of 3\%, 4\% and 5\% with the exchange interactions evaluated in the relaxed FE structures at each strain. The critical temperatures, determined from the peaks in the specific heat, are also indicated in the plots. (b) Magnetic critical temperatures as function of strain, obtained from MC simulations using exchange interactions from the centrosymmetric (solid lines and circles) and FE structures (dashed lines and crosses). Additionally, the mean field theory (MFT) results for the critical temperatures are shown in blue, with solid lines with triangles and dashed lines with squares for the centrosymmetric and FE structures, respectively. Experimental data from Maurel~\emph{et al.}~\cite{PhysRevB.92.024419} and Nakao~\emph{et al.}~\cite{NAKAO201418} are shown for comparison.}
	\label{fig.magMC_FE}
\end{figure}

The critical temperatures, determined from peaks in the specific heat, are indicated in the plots and also plotted separately as function of strain in Fig.~\ref{fig.magMC_FE}(b). The N\'eel temperature of 175~K for the unstrained structure can be compared to experimental values ranging from 227~K\cite{PhysRevB.92.024419} to 233~K~\cite{PhysRevB.64.134412}, indicating that the magnitude of the exchange interactions are somewhat underestimated in our calculations (even though further neighbor interactions or effects beyond the Heisenberg Hamiltonian can also contribute to these deviations). Calculations of the exchange interactions as functions of $U$ (see supplementary material~\cite{suppl}) reveal that the nearest neighbor interaction decreases in magnitude with increasing $U$. Thus, a somewhat smaller $U$ would result in a $T_\mathrm{N}$ in better agreement with experimental data and in the supplementary material it is estimated that $U=1.5~\mathrm{eV}$ would yield a $T_\mathrm{N}$ in agreement with experiment. However, to ensure comparability with previous studies~\cite{Becher2015,marthinsen_2016}, we did not adjust the $U$ value accordingly.

Fig.~\ref{fig.magMC_FE}(a), shows similar data as in Fig.~\ref{fig.magMC_strain} but using the exchange interactions of the non-centrosymmetric FE structures, i.e., corresponding to the dashed lines in Fig.~\ref{fig.Jofa}(a), for 3\%, 4\%, and 5\% strain. The data indicates a transition from G to C-type AFM, and then to FM order, again in agreement with the results from Fig.~\ref{fig.Eofa}. The critical temperatures are indicated and also plotted (dashed line, crosses), together with the data from Fig.~\ref{fig.magMC_strain} (solid line, circles) in Fig.~\ref{fig.magMC_FE}(b). In the range from 0-2\% strain the system exhibts G-AFM order and the  critical temperature decreases with strain. Keeping the centrosymmetric structure, the ordering temperature then increases again with strain, while the system makes the transition to C-AFM and then A-AFM order. On the other hand, considering the non-centrosymmetric FE structure, the critical temperature is enhanced, compared to the centrosymmetric case, at 3\% strain, and the G-AFM order is reinstated. Further increasing the strain decreases the critical temperature while the system switches to C-AFM and then FM order. This exemplifies the strong influence of the FE displacements on the magnetic properties of SrMnO$_3$. Not only does the FE order change the magnetic ground state, it also strongly affects the critical temperatures and thus the temperature dependence of the corresponding magnetic order parameters.

For comparison, mean field theory estimates of the critical temperatures are also included in Fig.~\ref{fig.magMC_FE} (blue triangles and squares for the centrosymmetric and FE structures, respectively). One can see that the mean field estimates follow exactly the same trend as the corresponding MC data, but, as expected, overestimate the MC data. Furthermore, experimental data~\cite{PhysRevB.92.024419,NAKAO201418} obtained for strained SrMnO$_3$ films is also included.
As already discussed, the MC results underestimate the experimental critical temperatures. Nevertheless, the qualitative trend, with critical temperatures that initially decrease, and then increase with strain, agrees very well with the available experimental data. Coincidentally, the mean field estimates agree very well with the experimental values and allow for a nice comparison of the strain dependence.

\section{Ferroelectricity}\label{sec.FE}

So far, we have fixed the polar FE distortion either to zero or to its relaxed zero temperature value, and studied the resulting differences in the temperature dependent magnetic properties.
In order to move towards a more comprehensive picture where FE and magnetic degrees of freedom are treated on equal footing, we now address the temperature dependence of the (strain dependent) FE order (while keeping the magnetic order fixed). To this end, we employ an effective Hamiltonian approach~\cite{PhysRevB.49.5828,PhysRevLett.73.1861,PhysRevB.52.6301}, which incorporates only the most important low energy structural distortions of the system, relative to an ideal, unstrained and centrosymmetric, cubic perovskite structure.
Thereby, the energy landscape is expressed in terms of gobal and local strain variables (long wavelength accoustic phonons) as well as local soft mode amplitudes (related to the polar phonon instability), while other structural distortions are not explicitly taken into account.

The required parameters for this effective Hamiltonian include parameters describing the soft mode self energy (the energy landscape for an isolated polar displacement), short range interaction parameters between the soft mode displacements in neighboring unit cells, the electronic part of the static dielectric constant and Born effective charges (which determine the long range dipole-dipole interactions), coupling parameters between the soft mode and strain, as well as the elastic constants. A proper definition of these parameters also requires specification of the corresponding local soft mode displacement vector. For a given material, all these parameters can be obtained from DFT calculations, as has been described in detail before~\cite{PhysRevB.49.5828,PhysRevB.52.6301,PhysRevB.82.134106}. Thus, the effective Hamiltonian approach enables a first-principles-based quantitative description of temperature dependent FE properties that does not involve any empirical parameters.

The parameterization employed here for SrMnO$_3$ roughly follows previously used schemes~\cite{PhysRevB.49.5828,PhysRevB.52.6301,PhysRevB.82.134106}  (note that the methods used to obtain the parameters differ somewhat in [\onlinecite{PhysRevB.49.5828},\onlinecite{PhysRevB.52.6301}] and [\onlinecite{PhysRevB.82.134106}]). The biggest difference to previous parameterization schemes stems from the fact that SrMnO$_3$ is not FE unless epitaxial strain is applied. In addition, we apply the effective Hamiltonian over a rather wide range of strain values.
In the following, we first discuss the necessary modifications to the parameterization scheme (Sec.~\ref{FE_par}), before the effective Hamiltonian, with the computed parameters, is used to study the FE properties of SrMnO$_3$ (Sec.~\ref{FE_res}). More details about how the various parameters are obtained using DFT calculations are contained in the supplementary material~\cite{suppl}.

\subsection{Determination of Parameters}\label{FE_par}

Usually, the local soft mode displacement vector can be determined, e.g., from the atomic displacements in the relaxed FE structure, or by identifying the dynamically unstable $\Gamma$-point phonon mode in the cubic structure. However, in the case of SrMnO$_3$ there is no structural instability in the unstrained cubic structure. We therefore consider the unstable phonon mode that develops under 3\,\% tensile strain. We find that this mode cannot directly be related to a particular phonon eigenmode of the cubic structure. Furthermore, due to the symmetry reduction in the strained tetragonal structure, the unstable phonon mode is a superposition of all 5 (three-fold degenerate) phonon modes of the cubic structure, not only of the 4 polar modes with $\Gamma_{15}$ symmetry~\cite{Dresselhaus}. In order to obtain a mode with the proper symmetry with respect to the cubic reference structure, we therefore project out the contribution corresponding to the non-polar cubic $\Gamma_{25}$ mode. This is discussed in more detail in the supplementary material~\cite{suppl}. The resulting soft mode displacements, $\xi_\mathrm{Sr}$, $\xi_\mathrm{Mn}$, $\xi_\mathrm{O_{||}}$ and $\xi_\mathrm{O_{\perp}}$ are listed in Table~\ref{tab_par}, with $\mathrm{O_{||}}$ and $\mathrm{O_{\perp}}$ respectively denoting O atoms located parallel or perpendicular to the displacement direction, relative to the Mn atoms.

With a choice for the FE soft mode, the remainder of the parametrisation can, in principle, be obtained according to the established schemes~\cite{PhysRevB.49.5828,PhysRevB.52.6301,PhysRevB.82.134106}. All parameters calculated in this work were determined by considering deviations around the unstrained cubic structure with G-type AFM order. Further details can be found in the supplementary material~\cite{suppl}. The resulting parameters are listed in Table~\ref{tab_par}, with a comparison to corresponding parameters for the prototypical FE BaTiO$_3$\cite{PhysRevB.82.134106}. The parameters include the cubic elastic constants $B_{11}$, $B_{12}$, and $B_{44}$, the strain-soft mode coupling parameters $B_\mathrm{1xx}$, $B_\mathrm{1yy}$, and $B_\mathrm{4yz}$, fourth and higher order self energy parameters $\alpha$, $\gamma$, and $k_i$, the mode effective mass $m^*$, Born effective charges $Z^*$ for each atom and the corresponding mode effective charge, the dielectric constant $\epsilon_\infty$, and the short range mode interaction parameters $j_k$, together with the second order self energy parameter $\kappa_2$. Here, $j_5$ and $j_7$ are set to zero, as has been motivated before~\cite{PhysRevB.52.6301} (they are expected to be small and require larger supercell calculations to be determined).

Most parameters are of comparable size and have the same sign as those for BaTiO$_3$. One notable exception is the short range coupling constant $j_2$ and some of the higher order self energy parameters $k_i$. It can also be noted that the strain-mode coupling parameters are somewhat stronger in SrMnO$_3$, as is favorable for the strain induced ferroelectricity. In agreement with previous work on CaMnO$_3$ and SrMnO$_3$~\cite{Bhattacharjee/Bousquet/Ghosez:2009,PhysRevB.79.205119,Ederer/Harris/Kovacik:2011}, the Born effective charge of the Mn cation, as well as $O_{||}$, are highly anomalous, in the sense that they are significantly larger then the corresponding formal charges. Accordingly, the soft mode displacement vector indicates a strong off-centering of the Mn cation with respect to the surrounding oxygen ligands, consistent with a $B$-site-driven FE distortion. 

\begin{table}[]
\centering
\caption{All parameters of the effective Hamiltonian for SrMnO$_3$. For comparison, an analogous parameterization for BaTiO$_3$\cite{PhysRevB.82.134106} is also given.}
\label{tab_par}
\begin{ruledtabular}
\begin{tabular}{p{3cm}l l  l }
Parameter & SrMnO$_3$ & BaTiO$_3$\cite{PhysRevB.82.134106} \\
\hline
$\xi_\mathrm{A}$             	  	& 0.039  			& 0.166  		\\
$\xi_\mathrm{B}$             	  	& 0.390       		& 0.770  		\\
$\xi_\mathrm{O_{||}}$        	  	& -0.666	   		& -0.55    		\\
$\xi_\mathrm{O_\perp}$         		& -0.449      		& -0.20    		\\
\hline
$a_0$ (\AA)              	 			& 3.79				& 3.99 			\\
$B_{11}$  (eV)          			& 114.485			& 126.73   		\\
$B_{12}$ (eV)           			& 35.436			& 41.76      	\\
$B_{44}$ (eV)           			& 42.519			& 49.24    		\\
$B_\mathrm{1xx}$ (eV/\AA$^2$)       & -214.736			& -185.35 	 	\\
$B_\mathrm{1yy}$ (eV/\AA$^2$)       & -10.540			& -3.2809  		\\
$B_\mathrm{4yz}$ (eV/\AA$^2$)       & -10.000    		& -14.550  		\\
$\alpha$ (eV/\AA$^4$)      	  		& 103.640			& 78.99       	\\
$\gamma$ (eV/\AA$^4$)       		& -224.286			& -115.48    	\\
$k_1$ (eV/\AA$^6$)       			& -928.579			& -267.98    	\\
$k_2$ (eV/\AA$^6$)       			& 1506.586			& -197.50    	\\
$k_3$ (eV/\AA$^6$)       			& 7712.958			& -830.20    	\\
$k_4$ (eV/\AA$^8$)       			& 4480.123			& 641.97    	\\
$m^*$                				& 22.042  			& 38.24 		\\
$Z^*_\mathrm{A}$   (e)          	& 2.576     		& 2.741 		\\
$Z^*_\mathrm{B}$  (e)          		& 7.673      		& 7.492 		\\
$Z^*_\mathrm{O_\perp}$ (e)       	& -1.717    		& -2.150 		\\
$Z^*_\mathrm{O_{||}}$ (e)   		& -6.813    		& -5.933 		\\
$Z^*$  (e)          				& 9.17  			& 10.33 		\\
\hline 
$\epsilon_\infty$  			  		& 10.68    			& 6.87 			\\
$\kappa_2$ (eV/\AA$^2$)			   	& 3.981	 			& 8.534 		\\
$j_1$ (eV/\AA$^2$)         	  		& -1.296  			& -2.084 		\\
$j_2$ (eV/\AA$^2$)         	  		& 4.347	 			& -1.129 		\\
$j_3$ (eV/\AA$^2$)         	  		& 0.272 			& 0.689 		\\
$j_4$ (eV/\AA$^2$)         	  		& -0.226  			& -0.611 		\\
$j_5$ (eV/\AA$^2$)         	  		& -	      			& - 			\\
$j_6$ (eV/\AA$^2$)         	  		& 0.100	    		& 0.277 		\\
$j_7$ (eV/\AA$^2$)         	  		& -            		& - 			\\
\end{tabular}
\end{ruledtabular}
\end{table}

Using the parameters as listed in Table~\ref{tab_par}, results in an ordering of local dipoles emerging at around 3\,\% strain, consistent with the DFT calculations. However, the emerging order is antiferroelectric rather than FE, in disagreement with expectations. Subsequent DFT supercell calculations confirmed that the expected FE arrangement of dipoles is indeed lower in energy than a potential antiferroelectric configuration at the given strain of 3\,\%, indicating a problem with the parameterisation, which, at low temperatures, should reproduce the DFT results. 

\begin{figure}[hbt!]
	\centering
%
%
\definecolor{mycolor1}{rgb}{0.00000,0.44700,0.74100}%
\definecolor{mycolor2}{rgb}{0.85000,0.32500,0.09800}%
\begin{tikzpicture}

\begin{axis}[%
width=0.4\textwidth,
height=0.2\textwidth,
at={(0\textwidth,0\textwidth)},
scale only axis,
xmin=0,
xmax=0.05,
xlabel style={font=\color{white!15!black}},
xlabel={Strain},
xtick = {0, 0.01, 0.02, 0.03, 0.04, 0.05},
x tick label style={/pgf/number format/fixed},
scaled ticks=false,
ymin=10,
ymax=16,
ylabel style={font=\color{white!15!black}},
ylabel={$\epsilon_\infty$},
axis background/.style={fill=white},
legend style={at={(0.451,0.7)}, anchor=south west, legend cell align=left, align=left, draw=white!15!black}
]
\addplot [color=mycolor1, line width=1.4pt, mark=x, mark options={solid, mycolor1}]
  table[row sep=crcr]{%
0	10.678578\\
0.01	11.288387\\
0.02	12.026865\\
0.03	12.974986\\
0.04	14.129367\\
0.05	15.663585\\
};
\addlegendentry{$\epsilon{}_{\text{xx}}$}

\addplot [color=mycolor2, line width=1.4pt, mark=o, mark options={solid, mycolor2}]
  table[row sep=crcr]{%
0	10.678578\\
0.01	10.554048\\
0.02	10.429379\\
0.03	10.474633\\
0.04	10.438982\\
0.05	10.488743\\
};
\addlegendentry{$\epsilon{}_{\text{zz}}$}

\node[right, align=left]
at (axis cs:0.0025,15.0) {(a)};
\end{axis}
\end{tikzpicture}%
 \\
	\input{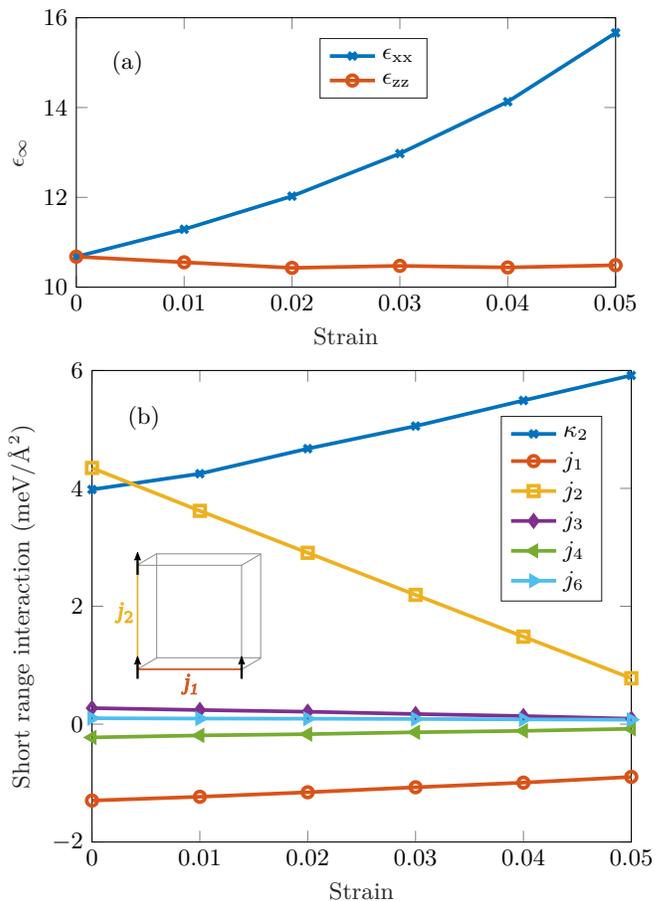}
	\caption{(a) Dielectric tensor components and (b) short range interactions, $j_k$, together with the quadratic self energy parameter, $\kappa_2$, as functions of strain. The inset in (b) illustrates the different nearest neighbor dipole couplings $j_1$ and $j_2$, respectively.}
	\label{fig.j_eps_strain}
\end{figure}

Within the effective Hamiltonian, the orientation of the dipoles relative to each other is determined by the short range couplings and the long range dipole-dipole interaction. 
To get further insight, we therefore calculate the dielectric tensor and the Born effective charges, which together determine the long range dipole-dipole interactions (proportional to $Z^{*2}/\epsilon_\infty$), as functions of strain. Note that in the cubic structure, the dielectric tensor has only one component, while under strain there is an in-plane component $\epsilon_{xx}=\epsilon_{yy}$, differing from the out-of-plane component $\epsilon_{zz}$.
We also calculate the strain dependence of the short range interactions, $j_k$, and of the second order self energy parameter $\kappa_2$, which are all evaluated from the same system of equations (see Eq. 15 in [\onlinecite{PhysRevB.82.134106}] or the supplementary material\cite{suppl}). The results are shown in Fig.~\ref{fig.j_eps_strain} (the Born effective charges were found to be insensitive to strain and are therefore not shown).

It can be seen that, while $\epsilon_{zz}$ is quite insensitive to strain, the in-plane dielectric constant $\epsilon_{xx}$ varies by nearly 50\% in the strain regime considered. Since, under tensile strain, the dipoles are expected to form in the $xy$-plane, the strain-dependent value of $\epsilon_{xx}$ is used as dielectric constant within the effective Hamiltonian and is also used to obtain the strain-dependent $j_k$ and $\kappa_2$. In addition, the short range interaction parameter $j_2$, which favors anti-parallel head-to-head arrangements of nearest neighbor dipoles when large and positive, strongly decreases in magnitude with strain.
Furthermore, also $\kappa_2$ shows a pronounced variation with strain. 

In the pioniering work by Zhong, Vanderbilt, and Rabe~\cite{PhysRevB.49.5828,PhysRevB.52.6301}, a coupling between strain and soft mode variables was only considered locally, i.e., corresponding to a modification of the quadratic part of the soft mode self energy that is linear in the strain variables, whereas the quadratic intersite couplings remain strain independent.
Our calculations clearly show that in the present case this is not a good approximation.
We note that the local soft mode-strain coupling that is already explicitly included in the effective Hamiltonian (resulting in an effective linear strain dependence of $\kappa_2$ described by the parameters $B_{1xx}$, $B_{1yy}$ and $B_{4yz}$),
has already been subtracted from the energies used to determine the parameters in Fig.~\ref{fig.j_eps_strain}(b). Thus, in principle $\kappa_2$ should not exhibit any strain dependence except for potential higher order contributions (at least quadratic in the strain).
The apparently linear strain dependence of $\kappa_2$ in Fig.~\ref{fig.j_eps_strain}(b) is due to the fact that we do not consider any strain dependence of the long range dipole-dipole interaction (except for the strain dependence of the dielectric constant), leading to an effective rescaling of the strain dependence of the other coefficients. A complete strain dependent description of dipole-dipole interactions is computationally challenging.
However, as a simple way to consider the dependence of the inter-site couplings on the global strain, the short range interaction parameters, including $\kappa_2$, are treated as strain dependent input parameters within our calculations. This is easily possible, since we always perform calculations at fixed global in-plane strain.

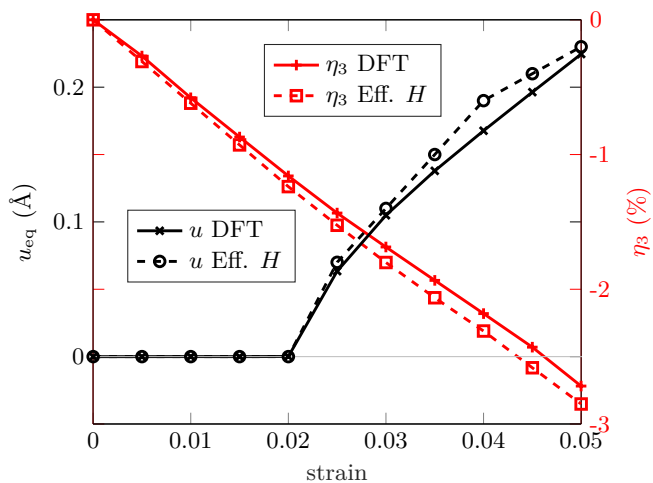
\begin{figure}[htbp]
	\centering
%
%
\begin{tikzpicture}

\pgfplotsset{set layers}

\begin{axis}[%
width=0.362\textwidth,
height=0.3\textwidth,
at={(0\textwidth,0\textwidth)},
scale only axis,
xmin=-1e-05,
xmax=0.0500001,
xlabel style={font=\color{white!15!black}},
xtick = {0, 0.01, 0.02, 0.03, 0.04, 0.05},
ytick = {-0.03,-0.02,-0.01,0},
yticklabels = {-3,-2,-1,0},
x tick label style={/pgf/number format/fixed},
y tick label style={/pgf/number format/fixed},
scaled ticks=false,
xlabel={strain},
separate axis lines,
every outer y axis line/.append style={red},
every y tick label/.append style={font=\color{red}},
every y tick/.append style={red},
ymin=-0.03,
ymax=0,
ylabel style={font=\color{red}},
ylabel={$\eta_{3}$ (\%)},
axis background/.style={fill=white},
yticklabel pos=right,
legend style={at={(0.35,0.76)}, anchor=south west, legend cell align=left, align=left, draw=white!15!black}
]

\addplot [color=red, line width=1.1pt, mark=+, mark options={solid, red}]
  table[row sep=crcr]{%
0	0\\
0.00499670402109431	-0.00268951878707968\\
0.00999340804218862	-0.00578773895847062\\
0.0150032959789057	-0.00870138431114043\\
0.02	-0.0115886618325644\\
0.0249967040210943	-0.0143441001977587\\
0.0299934080421886	-0.0168622280817403\\
0.0350032959789057	-0.019340804218853\\
0.04	-0.0218061964403428\\
0.0449967040210943	-0.0242979564930785\\
0.0499934080421884	-0.0271852340145023\\
};
\addlegendentry{$\eta_3$ DFT}

\addplot [color=red, dashed, line width=1.1pt, mark=square, mark options={solid, red}]
  table[row sep=crcr]{%
0	-0\\
0.005	-0.0030952116829381\\
0.01	-0.0061904233658762\\
0.015	-0.0092856350488143\\
0.02	-0.0123808467317524\\
0.025	-0.0152505067043776\\
0.03	-0.0180142954660396\\
0.035	-0.0206307851515789\\
0.04	-0.0230999757609956\\
0.045	-0.0258269397536269\\
0.05	-0.0285170789772276\\
};
\addlegendentry{$\eta_3$ Eff. $H$}

\end{axis}

\begin{axis}[%
width=0.362\textwidth,
height=0.3\textwidth,
at={(0\textwidth,0\textwidth)},
scale only axis,
xmin=-1e-05,
xmax=0.0500001,
xticklabels={,,},
xtick = {},
x tick label style={/pgf/number format/fixed},
scaled ticks=false,
xtick style={draw=none},
separate axis lines,
every outer y axis line/.append style={black},
every y tick label/.append style={font=\color{black}},
every y tick/.append style={black},
ymin=-0.05,
ymax=0.25,
ylabel style={font=\color{black}},
ylabel={\textcolor{black}{$u_{\mathrm{eq}}~(\mathrm{\AA})$}},
axis background/.style={fill=white},
yticklabel pos=left,
legend style={at={(0.07,0.35)}, anchor=south west, legend cell align=left, align=left, draw=white!15!black}
]
\addplot [color=black, line width=1.1pt, mark=x, mark options={solid, black}]
  table[row sep=crcr]{%
0	-3.49446165127484e-05\\
0.005	-2.04247802996863e-05\\
0.01	2.12204909524903e-05\\
0.015	-2.55419977785184e-05\\
0.02	0.000114930315006614\\
0.025	0.0634436618115094\\
0.03	0.10493641803305\\
0.035	0.137892836243664\\
0.04	0.167668530086164\\
0.045	0.196324908583579\\
0.05	0.224581007289913\\
};
\addlegendentry{$u$ DFT}

\addplot [color=black, dashed, line width=1.1pt, mark=o, mark options={solid, black}]
  table[row sep=crcr]{%
0	0\\
0.005	0\\
0.01	0\\
0.015	0\\
0.02	0\\
0.025	0.07\\
0.03	0.11\\
0.035	0.15\\
0.04	0.19\\
0.045	0.21\\
0.05	0.23\\
};
\addlegendentry{$u$ Eff. $H$}

\addplot[mark=none,color=gray!50] coordinates {(0,0) (1,0)};

\end{axis}
\end{tikzpicture}%
	\caption{Equilibrium FE soft mode amplitude, $u_{\mathrm{eq}}$, and out-of-plane strain component, $\eta_3$, obtained from the effective Hamiltonian and from DFT structural relaxations.}
	\label{fig.DFT_vs_MD}
\end{figure}

To double-check that the resulting parameterization of the effective Hamiltonian, with strain dependent parameters $j_k$, $\kappa_2$, and $\epsilon_{xx}$, is consistent with the DFT results over the whole considered strain range, Fig.~\ref{fig.DFT_vs_MD} shows the equilibrium soft mode displacement amplitude, $u_{\mathrm{eq}}$, and the out-of-plane strain, $\eta_3$, obtained by minimising the total energy described by the effective Hamiltonian, compared to the results of DFT structural relaxations. The energy of the effective Hamiltonian was minimised assuming a homogenous polarisation in the (110)-direction (the energy expression with further analysis is given in the supplementary material~\cite{suppl}). The agreement between the effective Hamiltonian and the DFT results is excellent for small strains, while the magnitude of $u_\mathrm{eq}$ and $\eta_3$ are slightly overestimated for larger strains. Nevertheless, the overall good agreement between the effective Hamiltonian and DFT results in Fig.~\ref{fig.DFT_vs_MD} corroborates that the effective Hamiltonian, with the given set of parameters, provides a consistent description of the strain-induced ferroelectricity in SrMnO$_3$.

\subsection{MD Simulations of the Effective Hamiltonian}\label{FE_res}

\begin{figure}[hbt!]
	\centering
	\input{Pol_of_strain_T.tex} \\
	{\includegraphics[width=0.273\textwidth]{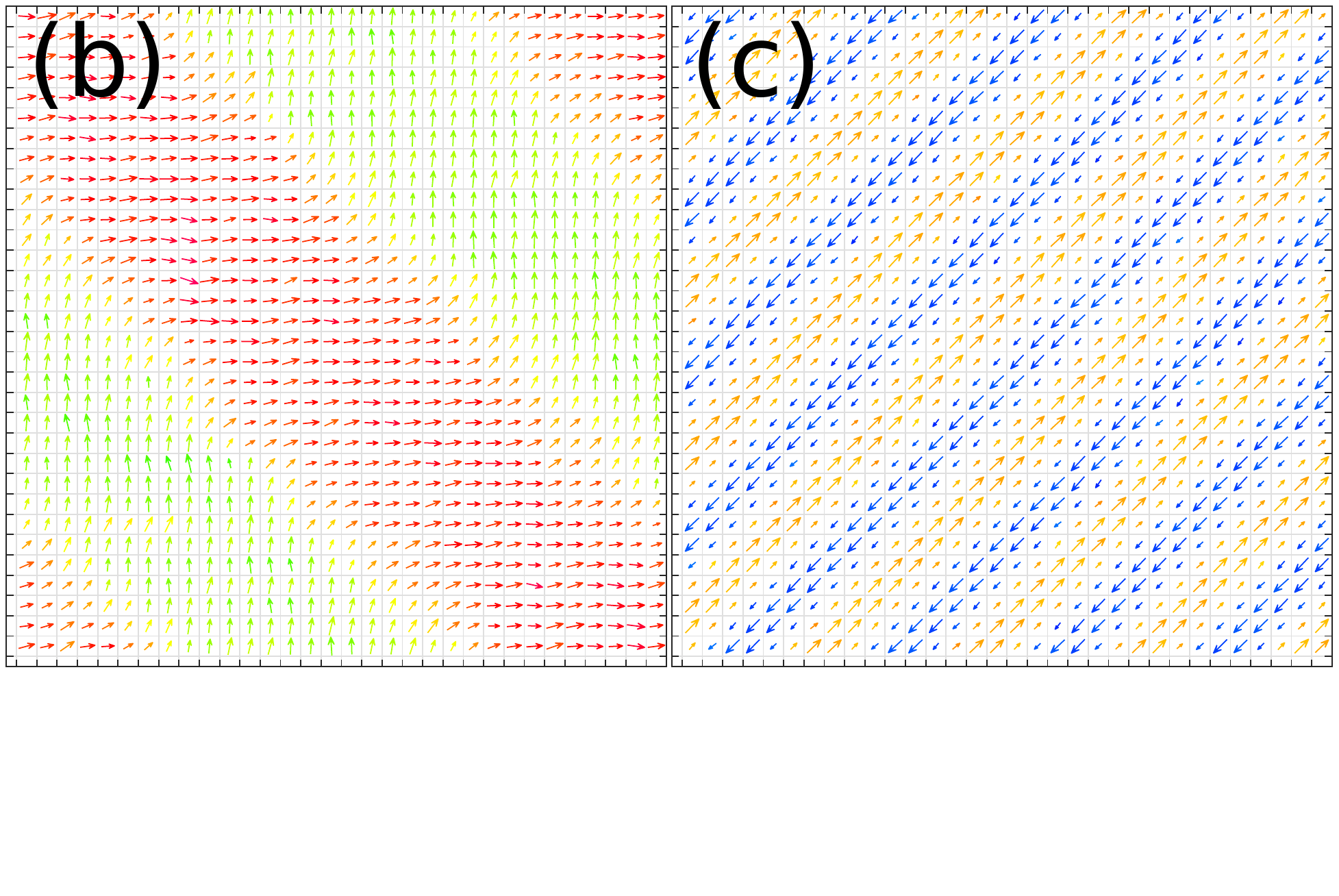}
%
%
\begin{tikzpicture}

\begin{axis}[%
width=0.13\textwidth,
height=0.13\textwidth,
at={(0\textwidth,0\textwidth)},
scale only axis,
xmin=0,
xmax=100,
xlabel style={font=\color{white!15!black}},
xlabel={$T$ (K)},
ymin=0,
ymax=1.5,
ylabel style={font=\color{white!15!black}},
ylabel={$\chi\text{ (arb. u.)}$},
axis background/.style={fill=white}
]
\addplot [color=black, line width=1.2pt, mark size=1.6pt, mark=o, mark options={solid, black}, forget plot]
  table[row sep=crcr]{%
100	0.2108646\\
95	0.225360526315789\\
90	0.244397111111111\\
85	0.246657764705882\\
80	0.26831075\\
75	0.3229864\\
70	0.311467571428571\\
65	0.388428\\
60	0.430782\\
55	0.533472\\
50	0.6245188\\
45	0.797379111111111\\
40	1.12115975\\
35	1.122624\\
30	0.8276792\\
25	0.6261732\\
20	0.5308612\\
15	0.4171994\\
10	0.3683708\\
5	0.3430942\\
};
\node[right, align=left]
at (axis cs:40,1.25) {40 K};

\node[right, align=left]
at (axis cs:1,1.2) {(d)};

\end{axis}
\end{tikzpicture}%
}
	\caption{(a) Polarization as function of temperature, simulated upon heating, for various tensile strains in SrMnO$_3$. Snapshots from MD simulations, with arrows indicating $u$ in the $xy$-plane, averaged along the $z$-direction, for (b) $\eta=0.04$, $T=270~\mathrm{K}$ and (c) $\eta=0.025$, $T=5~\mathrm{K}$. The colorwheel illustrates the directions of the color coded arrows in (b)-(c). (d) Susceptibility $\chi$ as function of temperature for 2.5\% strain. }
	\label{fig.pol_strain_T}
\end{figure}

Using the parameterization discussed in the previous section, we perform MD simulations for various strains in the range of 0-5\%. In each case, a simulated heating is performed, where a homogeneously polarized FE state is initialized, and the simulation is first performed at a low temperature, which is then increased, in increments of 5 K, using the final configuration of the previous simulation as initialization for the next one. The resulting polarizations as function of temperature, for different strains, are shown in Fig.~\ref{fig.pol_strain_T}(a).
For small strains, $\eta<3$\,\%, the initialized FE state is unstable, i.e., it vanishes during the thermalization phase, and no spontaneous polarization occurs at finite temperature. For $\eta \geq 3\%$ the polarization remains stable at low temperatures and then drops to zero at the strain dependent FE Curie temperature. Both the saturation polarization as well as the Curie temperature increase strongly with strain. Around 4\,\% strain, the Curie temperature exceeds room temperature, and the saturation polarization is $\sim$\,50\,$\mu$C/cm$^2$, which corresponds to about twice that of bulk BaTiO$_3$. Very recent experiments on highly strained SrMnO$_3$ films showed a remnant polarization of $55~\mathrm{\mu C/cm^2}$ at 3.8\% strain\cite{PhysRevB.97.235135}, which is of similar size as the polarizations found here.

In the large strain regime of $\eta \geq 4\%$, an additional feature appears, corresponding to a sharp drop in $P(T)$, indicating a transition between different polar phases. As illustrated in Fig.~\ref{fig.pol_strain_T}(b) for $T=270$\,K and $\eta=0.04$, a domain structure appears below $T_\mathrm{C}$, with local polarization along either [100] or [010], separated by $90^\circ$ domain walls, and resulting in an effective global polarization along [110]. At lower temperatures, between around 200\,K and 300 \,K, the system then transforms into the uniform FE state with polarization along the [110]-direction, leading to the observed sharp change in total polarization. Similar behavior has been observed previously in simulations for epitaxially strained BaTiO$_3$\cite{doi:10.1063/1.4930306} under tensile strain. The 90$^\circ$ domain state also occurs in cooling simulations starting from temperatures above $T_\mathrm{C}$, and for the largest strain of 5\% seems to persist even down to the lowest temperatures. This indicates a rather rich phase diagram with potentially coexisting phases and different competing domain configurations. Here we want to focus only on the main features, in particular on the onset of polar order under tensile strain, and we thus leave a more detailed investigation of this phase diagram for future work. 

From Fig.~\ref{fig.pol_strain_T}(a) it can also be seen that no spontaneous polarization appears at 2.5\,\% strain, in contrast to the results shown in Fig.~\ref{fig.DFT_vs_MD}. However, the temperature dependence of the electric susceptibility $\chi=\frac{<P^2>-<P>^2}{T}$ at this strain, illustrated in Fig.~\ref{fig.pol_strain_T}(d), reveals an anomaly with a clear maximum at 40 K, indicating a phase transition. 
Fig.~\ref{fig.pol_strain_T}(c) shows a snapshot of the corresponding dipole configuration at 5\,K in the $x$-$y$-plane, averaged over $z$. It can be seen that the dipoles form antiparallel stripe domains, with zero net polarization. Such an inhomogeneous configuration is obviously not captured by the results shown in Fig.~\ref{fig.DFT_vs_MD}, where a homogeneous polarization along the [110]-direction is assumed.
Thus, at 2.5\,\% strain, the 180$^\circ$ stripe domain state appears to be more stable than the homogeneous FE state.
We note that the results in Figs.~\ref{fig.pol_strain_T}(c)-(d) were obtained from a simulated cooling, rather than heating, since extremely long equilibration times are otherwise needed at low temperature for the system to turn from the homogenously polarized initial state to the stripe domain state. Again, the apparance of such domain states indicates a more complex phase diagram, the full exploration of which, however, is beyond the scope of the current work. Furthermore, the existence of these domains still needs to be established experimentally, and further DFT studies could also be beneficial in future work. 

\section{Discussion of the Strain-Temperature Multiferroic Phase Diagram}\label{sec.PD}

\begin{figure}[hbt!]
	\centering
	\includegraphics[width=0.494\textwidth]{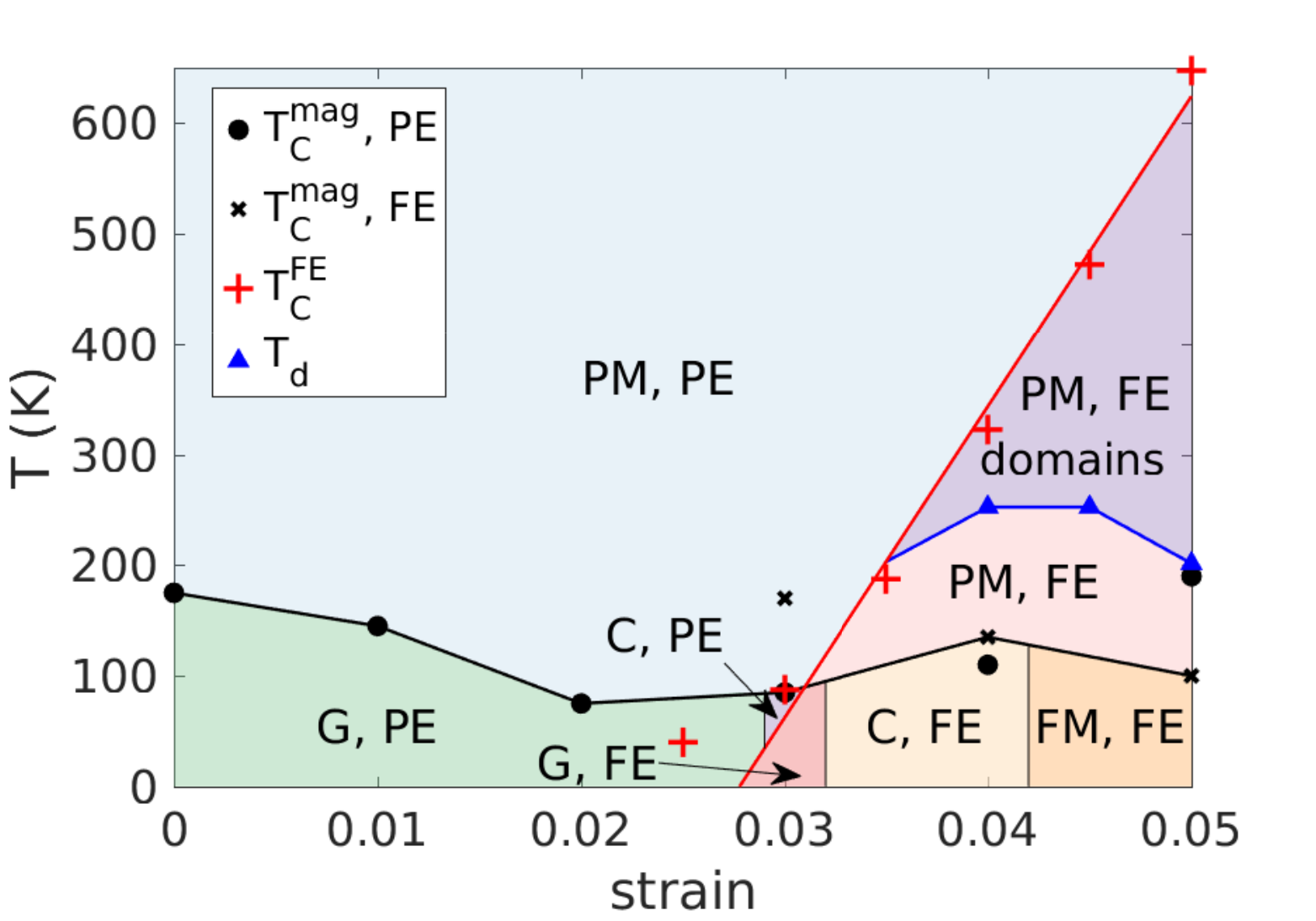}
	\caption{Strain-temperature ferroic phase diagram of SrMnO$_3$. Black circles show magnetic critical temperatures calculated using exchange interactions obtained for the centrosymmetric (PE) structures, while the black crosses are calculated using exchange interactions obtained for the FE structures. Red plus signs indicate the calculated FE critical temperatures and the red line is a linear fit to these. The blue triangles show the temperatures above which FE domains form at large strains. The positions of vertical lines separating magnetic regions have been estimated from the data in Fig.~\ref{fig.Eofa} and are mainly a guide to the eye.}
	\label{fig.phasediag}
\end{figure}

The preceding sections presented the individually calculated magnetic and ferroelectric phase diagrams as function of strain and temperature. By combining these results, a complete multiferroic strain-temperature phase diagram of SrMnO$_3$ is obtained, which is summarized in Fig.~\ref{fig.phasediag}.
As follows from the results presented in Sec.~\ref{FE_res}, the FE critical temperature increases nearly linearly with strain after its onset at around 3\.\%. Note that the red line in Fig.~\ref{fig.phasediag} shows a linear fit of $T_\mathrm{C}^\mathrm{FE}$ as function of strain, excluding the point at 2.5\% strain, which falls away from this trend and corresponds to the appearance of the 180$^\circ$ stripe domain state (see Fig.~\ref{fig.pol_strain_T}(c)). Compared to $T_\mathrm{C}^\mathrm{FE}$, the previously discussed strain dependence of the magnetic critical temperature is much more moderate. Consequently, for strains slighly above 3\,\%, the FE and magnetic critical temperatures cross. In the strain region around this crossing, both critical temperature are close to each other and pronounced magnetoelectric coupling effects can be expected, such as, e.g., thermally mediated magnetoelectric coupling~\cite{PhysRevLett.114.177205} or multicaloric effects~\cite{planes_multical}.

Furthermore, as already discussed in Sec.~\ref{mag_MCres}, the magnetic critical temperatures obtained for the FE and the  centrosymmetric structures, respectively, are vastly different, and also correspond to different magnetic orders. This suggests that an applied electric field might significantly alter the magnetic critical temperature and potentially even induce transitions between different magnetic states.

Judging from the linear fit of $T_\mathrm{C}^\mathrm{FE}$, a small C-AFM paramagnetic region appears at approximately 3\% strain and temperatures above $T_\mathrm{C}^\mathrm{FE}$, whereas in the corresponding FE low temperature phase, the magnetic ground state is G-AFM (with a significantly higher hypothetical magnetic critical temperature). Thus, one can expect a magnetic transition to occur in that strain regime, related to the decrease in FE spontaneous polarization with increasing temperature, potentially also resulting in a strong electric-field dependence of the magnetic order.
Apart from the region around 3\% strain, another very interesting feature of this rich phase diagram is the multiferroic region with simultaneous FM and FE order occuring for strains above $\sim$\,4\,\% and temperatures below $\sim$\,100~K.

Of course, due to the systematic uncertainties involved in the parameterization of both the magnetic Heisenberg Hamiltonian and the FE effective Hamiltonian, the exact phase boundaries in this system remain to be experimentally verified.
On the computational side, a more explicit treatment of the coupling between the structural and magnetic degrees of freedom, and allowing for simultaneous fluctuations in both quantities, would be desirable.
Nevertheless, the possible existence of a region with strongly coupled magnetic and FE order is highly promising, as it indicates to possibility to control magnetic order using an electric field.

\section{Summary and Conclusions}\label{sec.concl}

In this work, DFT calculations have been used to construct effective Hamiltonians for both the magnetic and FE structural degrees of freedom. The resulting temperature and strain dependent ferroic phase diagram has then been obtained from MC and MD simulations. The coupling between magnetism and FE order has been incorporated by considering the change in the Heisenberg exchange interactions, and the corresponding changes to the magnetic phase diagram, due to the FE structural distortions.
The resulting phase diagram (Fig.~\ref{fig.phasediag}) exhibits various regions of particular interest for further investigation of magnetoelectric coupling phenomena, such as the regime around or just above 3\% tensile strain, where the magnetic and FE critical temperatures nearly coincide, or the FM-FE region at higher strain.
Both regions can be technologically interesting, as they potentially allow for electric field control of magnetic properties.

Compared to currently available experimental data~\cite{PhysRevB.64.134412}, the calculations presented here underestimate the magnetic critical temperature of bulk SrMnO$_3$ by about 60~K. This is probably related to the strong sensitivity of the calculated exchange interactions on the specific value of the Hubbard $U$ used in the DFT calculations (see supplementary material~\cite{suppl}), but can also, at least partly, be caused by a non-Heisenberg character of the magnetic interactions in this system.
Furthermore, experiments have indicated the appearance of a polar phase at a tensile strain of 1.7\%~\cite{Becher2015}, somewhat lower than the critical strain of about 2.5\% found here. We note that the use of a different exchange-correlation functional, such as, e.g., the PBE functional used in Ref.~\onlinecite{PhysRevLett.104.207204}, leads to changes in the equilibrium lattice parameters of that order of magnitude, which is within the typical limitations of current DFT functionals.
Despite these quantitative differences, the qualitative trends as well as the main structure of the phase diagram are expected to be well captured by the computational methods applied here 

Although we have studied the effect of strain on SrMnO$_3$, it is known that similar effects can be achieved by Ba-substitution~\cite{PhysRevLett.107.137601}. Thus, for experimental realisations of some of the effects discussed here, a combination of Ba-substitution and strain might be the most promising path. Furthermore, it has been suggested that strain results in increased oxygen vacancy concentrations, which supress ferroelectricity~\cite{marthinsen_2016}. Hence, it might also be relevant to take into consideration the effects of oxygen vacancies in future studies. Finally, as the perhaps most important continuation for future work, it would be desirable to develop more refined models for the coupling between the ferroic degrees of freedom, treating magnetism and FE on equal footing.

\section{Acknowledgments}
We are grateful to Anna Gr\"unebohm for discussions regarding the parameterisation and use of the FE effective Hamiltonian. This work was supported by the Swiss National Science Foundation (project code 200021E-162297) and the German Science Foundation under the priority program SPP 1599 (``Ferroic Cooling''). Computational work was performed on resources provided by the Swiss National Supercomputing Centre (CSCS).

\bibliography{literature}{}

\begin{thebibliography}{57}%
\makeatletter
\providecommand \@ifxundefined [1]{%
 \@ifx{#1\undefined}
}%
\providecommand \@ifnum [1]{%
 \ifnum #1\expandafter \@firstoftwo
 \else \expandafter \@secondoftwo
 \fi
}%
\providecommand \@ifx [1]{%
 \ifx #1\expandafter \@firstoftwo
 \else \expandafter \@secondoftwo
 \fi
}%
\providecommand \natexlab [1]{#1}%
\providecommand \enquote  [1]{``#1''}%
\providecommand \bibnamefont  [1]{#1}%
\providecommand \bibfnamefont [1]{#1}%
\providecommand \citenamefont [1]{#1}%
\providecommand \href@noop [0]{\@secondoftwo}%
\providecommand \href [0]{\begingroup \@sanitize@url \@href}%
\providecommand \@href[1]{\@@startlink{#1}\@@href}%
\providecommand \@@href[1]{\endgroup#1\@@endlink}%
\providecommand \@sanitize@url [0]{\catcode `\\12\catcode `\$12\catcode
  `\&12\catcode `\#12\catcode `\^12\catcode `\_12\catcode `\%12\relax}%
\providecommand \@@startlink[1]{}%
\providecommand \@@endlink[0]{}%
\providecommand \url  [0]{\begingroup\@sanitize@url \@url }%
\providecommand \@url [1]{\endgroup\@href {#1}{\urlprefix }}%
\providecommand \urlprefix  [0]{URL }%
\providecommand \Eprint [0]{\href }%
\providecommand \doibase [0]{http://dx.doi.org/}%
\providecommand \selectlanguage [0]{\@gobble}%
\providecommand \bibinfo  [0]{\@secondoftwo}%
\providecommand \bibfield  [0]{\@secondoftwo}%
\providecommand \translation [1]{[#1]}%
\providecommand \BibitemOpen [0]{}%
\providecommand \bibitemStop [0]{}%
\providecommand \bibitemNoStop [0]{.\EOS\space}%
\providecommand \EOS [0]{\spacefactor3000\relax}%
\providecommand \BibitemShut  [1]{\csname bibitem#1\endcsname}%
\let\auto@bib@innerbib\@empty
\bibitem [{\citenamefont {Ramesh}\ and\ \citenamefont
  {Spaldin}(2007)}]{Ramesh2007}%
  \BibitemOpen
  \bibfield  {author} {\bibinfo {author} {\bibfnamefont {R.}~\bibnamefont
  {Ramesh}}\ and\ \bibinfo {author} {\bibfnamefont {N.~A.}\ \bibnamefont
  {Spaldin}},\ }\href {http://dx.doi.org/10.1038/nmat1805} {\bibfield
  {journal} {\bibinfo  {journal} {Nature Materials}\ }\textbf {\bibinfo
  {volume} {6}} (\bibinfo {year} {2007})}\BibitemShut {NoStop}%
\bibitem [{\citenamefont {Spaldin}(2017)}]{Spaldin2017}%
  \BibitemOpen
  \bibfield  {author} {\bibinfo {author} {\bibfnamefont {N.~A.}\ \bibnamefont
  {Spaldin}},\ }\href {http://dx.doi.org/10.1038/natrevmats.2017.17} {\bibfield
   {journal} {\bibinfo  {journal} {Nature Reviews Materials}\ }\textbf
  {\bibinfo {volume} {2}},\ \bibinfo {pages} {17017} (\bibinfo {year}
  {2017})}\BibitemShut {NoStop}%
\bibitem [{\citenamefont {Khomskii}(2009)}]{Khomskii:2009}%
  \BibitemOpen
  \bibfield  {author} {\bibinfo {author} {\bibfnamefont {D.}~\bibnamefont
  {Khomskii}},\ }\href {\doibase 10.1103/Physics.2.20} {\bibfield  {journal}
  {\bibinfo  {journal} {Physics}\ }\textbf {\bibinfo {volume} {2}},\ \bibinfo
  {pages} {20} (\bibinfo {year} {2009})}\BibitemShut {NoStop}%
\bibitem [{\citenamefont {Chang}\ \emph {et~al.}(2015)\citenamefont {Chang},
  \citenamefont {Mani}, \citenamefont {Lisenkov},\ and\ \citenamefont
  {Ponomareva}}]{PhysRevLett.114.177205}%
  \BibitemOpen
  \bibfield  {author} {\bibinfo {author} {\bibfnamefont {C.-M.}\ \bibnamefont
  {Chang}}, \bibinfo {author} {\bibfnamefont {B.~K.}\ \bibnamefont {Mani}},
  \bibinfo {author} {\bibfnamefont {S.}~\bibnamefont {Lisenkov}}, \ and\
  \bibinfo {author} {\bibfnamefont {I.}~\bibnamefont {Ponomareva}},\ }\href
  {\doibase 10.1103/PhysRevLett.114.177205} {\bibfield  {journal} {\bibinfo
  {journal} {Phys. Rev. Lett.}\ }\textbf {\bibinfo {volume} {114}},\ \bibinfo
  {pages} {177205} (\bibinfo {year} {2015})}\BibitemShut {NoStop}%
\bibitem [{\citenamefont {Planes}\ \emph {et~al.}(2014)\citenamefont {Planes},
  \citenamefont {Castan},\ and\ \citenamefont {Saxena}}]{planes_multical}%
  \BibitemOpen
  \bibfield  {author} {\bibinfo {author} {\bibfnamefont {A.}~\bibnamefont
  {Planes}}, \bibinfo {author} {\bibfnamefont {T.}~\bibnamefont {Castan}}, \
  and\ \bibinfo {author} {\bibfnamefont {A.}~\bibnamefont {Saxena}},\ }\href
  {\doibase 10.1080/14786435.2014.899438} {\bibfield  {journal} {\bibinfo
  {journal} {Philosophical Magazine}\ }\textbf {\bibinfo {volume} {94}},\
  \bibinfo {pages} {1893} (\bibinfo {year} {2014})}\BibitemShut {NoStop}%
\bibitem [{\citenamefont {Lee}\ and\ \citenamefont
  {Rabe}(2010)}]{PhysRevLett.104.207204}%
  \BibitemOpen
  \bibfield  {author} {\bibinfo {author} {\bibfnamefont {J.~H.}\ \bibnamefont
  {Lee}}\ and\ \bibinfo {author} {\bibfnamefont {K.~M.}\ \bibnamefont {Rabe}},\
  }\href {\doibase 10.1103/PhysRevLett.104.207204} {\bibfield  {journal}
  {\bibinfo  {journal} {Phys. Rev. Lett.}\ }\textbf {\bibinfo {volume} {104}},\
  \bibinfo {pages} {207204} (\bibinfo {year} {2010})}\BibitemShut {NoStop}%
\bibitem [{\citenamefont {Becher}\ \emph {et~al.}(2015)\citenamefont {Becher},
  \citenamefont {Maurel}, \citenamefont {Aschauer}, \citenamefont {Lilienblum},
  \citenamefont {Mag\'{e}}, \citenamefont {Meier}, \citenamefont {Langenberg},
  \citenamefont {Trassin}, \citenamefont {Blasco}, \citenamefont {Krug},
  \citenamefont {Algarabel}, \citenamefont {Spaldin}, \citenamefont {Pardo},\
  and\ \citenamefont {Fiebig}}]{Becher2015}%
  \BibitemOpen
  \bibfield  {author} {\bibinfo {author} {\bibfnamefont {C.}~\bibnamefont
  {Becher}}, \bibinfo {author} {\bibfnamefont {L.}~\bibnamefont {Maurel}},
  \bibinfo {author} {\bibfnamefont {U.}~\bibnamefont {Aschauer}}, \bibinfo
  {author} {\bibfnamefont {M.}~\bibnamefont {Lilienblum}}, \bibinfo {author}
  {\bibfnamefont {C.}~\bibnamefont {Mag\'{e}}}, \bibinfo {author}
  {\bibfnamefont {D.}~\bibnamefont {Meier}}, \bibinfo {author} {\bibfnamefont
  {E.}~\bibnamefont {Langenberg}}, \bibinfo {author} {\bibfnamefont
  {M.}~\bibnamefont {Trassin}}, \bibinfo {author} {\bibfnamefont
  {J.}~\bibnamefont {Blasco}}, \bibinfo {author} {\bibfnamefont {I.~P.}\
  \bibnamefont {Krug}}, \bibinfo {author} {\bibfnamefont {P.~A.}\ \bibnamefont
  {Algarabel}}, \bibinfo {author} {\bibfnamefont {N.~A.}\ \bibnamefont
  {Spaldin}}, \bibinfo {author} {\bibfnamefont {J.~A.}\ \bibnamefont {Pardo}},
  \ and\ \bibinfo {author} {\bibfnamefont {M.}~\bibnamefont {Fiebig}},\ }\href
  {http://dx.doi.org/10.1038/nnano.2015.108} {\bibfield  {journal} {\bibinfo
  {journal} {Nature Nanotechnology}\ }\textbf {\bibinfo {volume} {10}},\
  \bibinfo {pages} {661} (\bibinfo {year} {2015})},\ \bibinfo {note}
  {letter}\BibitemShut {NoStop}%
\bibitem [{\citenamefont {Guzman}\ \emph {et~al.}(2016)\citenamefont {Guzman},
  \citenamefont {Maurel}, \citenamefont {Langenberg}, \citenamefont {Lupini},
  \citenamefont {Algarabel}, \citenamefont {Pardo},\ and\ \citenamefont
  {Magen}}]{acs.nanolett.5b04455}%
  \BibitemOpen
  \bibfield  {author} {\bibinfo {author} {\bibfnamefont {R.}~\bibnamefont
  {Guzman}}, \bibinfo {author} {\bibfnamefont {L.}~\bibnamefont {Maurel}},
  \bibinfo {author} {\bibfnamefont {E.}~\bibnamefont {Langenberg}}, \bibinfo
  {author} {\bibfnamefont {A.~R.}\ \bibnamefont {Lupini}}, \bibinfo {author}
  {\bibfnamefont {P.~A.}\ \bibnamefont {Algarabel}}, \bibinfo {author}
  {\bibfnamefont {J.~A.}\ \bibnamefont {Pardo}}, \ and\ \bibinfo {author}
  {\bibfnamefont {C.}~\bibnamefont {Magen}},\ }\href {\doibase
  10.1021/acs.nanolett.5b04455} {\bibfield  {journal} {\bibinfo  {journal}
  {Nano Letters}\ }\textbf {\bibinfo {volume} {16}},\ \bibinfo {pages} {2221}
  (\bibinfo {year} {2016})}\BibitemShut {NoStop}%
\bibitem [{\citenamefont {Langenberg}\ \emph {et~al.}(2017)\citenamefont
  {Langenberg}, \citenamefont {Maurel}, \citenamefont {Marcano}, \citenamefont
  {Guzman}, \citenamefont {Strichovanec}, \citenamefont {Prokscha},
  \citenamefont {Magen}, \citenamefont {Algarabel},\ and\ \citenamefont
  {Pardo}}]{ADMI201601040}%
  \BibitemOpen
  \bibfield  {author} {\bibinfo {author} {\bibfnamefont {E.}~\bibnamefont
  {Langenberg}}, \bibinfo {author} {\bibfnamefont {L.}~\bibnamefont {Maurel}},
  \bibinfo {author} {\bibfnamefont {N.}~\bibnamefont {Marcano}}, \bibinfo
  {author} {\bibfnamefont {R.}~\bibnamefont {Guzman}}, \bibinfo {author}
  {\bibfnamefont {P.}~\bibnamefont {Strichovanec}}, \bibinfo {author}
  {\bibfnamefont {T.}~\bibnamefont {Prokscha}}, \bibinfo {author}
  {\bibfnamefont {C.}~\bibnamefont {Magen}}, \bibinfo {author} {\bibfnamefont
  {P.~A.}\ \bibnamefont {Algarabel}}, \ and\ \bibinfo {author} {\bibfnamefont
  {J.~A.}\ \bibnamefont {Pardo}},\ }\href {\doibase 10.1002/admi.201601040}
  {\bibfield  {journal} {\bibinfo  {journal} {Advanced Materials Interfaces}\
  }\textbf {\bibinfo {volume} {4}},\ \bibinfo {pages} {1601040} (\bibinfo
  {year} {2017})},\ \bibinfo {note} {1601040}\BibitemShut {NoStop}%
\bibitem [{\citenamefont {Guo}\ \emph {et~al.}(2018)\citenamefont {Guo},
  \citenamefont {Wang}, \citenamefont {Yuan}, \citenamefont {He}, \citenamefont
  {Lu}, \citenamefont {Chen}, \citenamefont {Yang}, \citenamefont {Wang},
  \citenamefont {Erni}, \citenamefont {Rossell}, \citenamefont {Gopalan},
  \citenamefont {Xiang}, \citenamefont {Tokura},\ and\ \citenamefont
  {Yu}}]{PhysRevB.97.235135}%
  \BibitemOpen
  \bibfield  {author} {\bibinfo {author} {\bibfnamefont {J.~W.}\ \bibnamefont
  {Guo}}, \bibinfo {author} {\bibfnamefont {P.~S.}\ \bibnamefont {Wang}},
  \bibinfo {author} {\bibfnamefont {Y.}~\bibnamefont {Yuan}}, \bibinfo {author}
  {\bibfnamefont {Q.}~\bibnamefont {He}}, \bibinfo {author} {\bibfnamefont
  {J.~L.}\ \bibnamefont {Lu}}, \bibinfo {author} {\bibfnamefont {T.~Z.}\
  \bibnamefont {Chen}}, \bibinfo {author} {\bibfnamefont {S.~Z.}\ \bibnamefont
  {Yang}}, \bibinfo {author} {\bibfnamefont {Y.~J.}\ \bibnamefont {Wang}},
  \bibinfo {author} {\bibfnamefont {R.}~\bibnamefont {Erni}}, \bibinfo {author}
  {\bibfnamefont {M.~D.}\ \bibnamefont {Rossell}}, \bibinfo {author}
  {\bibfnamefont {V.}~\bibnamefont {Gopalan}}, \bibinfo {author} {\bibfnamefont
  {H.~J.}\ \bibnamefont {Xiang}}, \bibinfo {author} {\bibfnamefont
  {Y.}~\bibnamefont {Tokura}}, \ and\ \bibinfo {author} {\bibfnamefont
  {P.}~\bibnamefont {Yu}},\ }\href {\doibase 10.1103/PhysRevB.97.235135}
  {\bibfield  {journal} {\bibinfo  {journal} {Phys. Rev. B}\ }\textbf {\bibinfo
  {volume} {97}},\ \bibinfo {pages} {235135} (\bibinfo {year}
  {2018})}\BibitemShut {NoStop}%
\bibitem [{\citenamefont {Wang}\ \emph {et~al.}(2016)\citenamefont {Wang},
  \citenamefont {Zhang}, \citenamefont {Bai}, \citenamefont {Liu},
  \citenamefont {Zhang}, \citenamefont {Wang}, \citenamefont {Li},
  \citenamefont {Ma}, \citenamefont {Zhao}, \citenamefont {Sun}, \citenamefont
  {Wang}, \citenamefont {Wang},\ and\ \citenamefont
  {Zhang}}]{doi:10.1063/1.4960463}%
  \BibitemOpen
  \bibfield  {author} {\bibinfo {author} {\bibfnamefont {F.}~\bibnamefont
  {Wang}}, \bibinfo {author} {\bibfnamefont {Y.~Q.}\ \bibnamefont {Zhang}},
  \bibinfo {author} {\bibfnamefont {Y.}~\bibnamefont {Bai}}, \bibinfo {author}
  {\bibfnamefont {W.}~\bibnamefont {Liu}}, \bibinfo {author} {\bibfnamefont
  {H.~R.}\ \bibnamefont {Zhang}}, \bibinfo {author} {\bibfnamefont {W.~Y.}\
  \bibnamefont {Wang}}, \bibinfo {author} {\bibfnamefont {S.~K.}\ \bibnamefont
  {Li}}, \bibinfo {author} {\bibfnamefont {S.}~\bibnamefont {Ma}}, \bibinfo
  {author} {\bibfnamefont {X.~G.}\ \bibnamefont {Zhao}}, \bibinfo {author}
  {\bibfnamefont {J.~R.}\ \bibnamefont {Sun}}, \bibinfo {author} {\bibfnamefont
  {Z.~H.}\ \bibnamefont {Wang}}, \bibinfo {author} {\bibfnamefont {Z.~J.}\
  \bibnamefont {Wang}}, \ and\ \bibinfo {author} {\bibfnamefont {Z.~D.}\
  \bibnamefont {Zhang}},\ }\href {\doibase 10.1063/1.4960463} {\bibfield
  {journal} {\bibinfo  {journal} {Applied Physics Letters}\ }\textbf {\bibinfo
  {volume} {109}},\ \bibinfo {pages} {052403} (\bibinfo {year} {2016})},\
  \Eprint {http://arxiv.org/abs/https://doi.org/10.1063/1.4960463}
  {https://doi.org/10.1063/1.4960463} \BibitemShut {NoStop}%
\bibitem [{\citenamefont {Bai}\ \emph {et~al.}(2017)\citenamefont {Bai},
  \citenamefont {Yang}, \citenamefont {Dong}, \citenamefont {Zhang},
  \citenamefont {Bai},\ and\ \citenamefont {Tang}}]{BAI201757}%
  \BibitemOpen
  \bibfield  {author} {\bibinfo {author} {\bibfnamefont {J.}~\bibnamefont
  {Bai}}, \bibinfo {author} {\bibfnamefont {J.}~\bibnamefont {Yang}}, \bibinfo
  {author} {\bibfnamefont {W.}~\bibnamefont {Dong}}, \bibinfo {author}
  {\bibfnamefont {Y.}~\bibnamefont {Zhang}}, \bibinfo {author} {\bibfnamefont
  {W.}~\bibnamefont {Bai}}, \ and\ \bibinfo {author} {\bibfnamefont
  {X.}~\bibnamefont {Tang}},\ }\href {\doibase
  https://doi.org/10.1016/j.tsf.2017.08.052} {\bibfield  {journal} {\bibinfo
  {journal} {Thin Solid Films}\ }\textbf {\bibinfo {volume} {644}},\ \bibinfo
  {pages} {57 } (\bibinfo {year} {2017})}\BibitemShut {NoStop}%
\bibitem [{\citenamefont {Marthinsen}\ \emph {et~al.}(2016)\citenamefont
  {Marthinsen}, \citenamefont {Faber}, \citenamefont {Aschauer}, \citenamefont
  {Spaldin},\ and\ \citenamefont {Selbach}}]{marthinsen_2016}%
  \BibitemOpen
  \bibfield  {author} {\bibinfo {author} {\bibfnamefont {A.}~\bibnamefont
  {Marthinsen}}, \bibinfo {author} {\bibfnamefont {C.}~\bibnamefont {Faber}},
  \bibinfo {author} {\bibfnamefont {U.}~\bibnamefont {Aschauer}}, \bibinfo
  {author} {\bibfnamefont {N.~A.}\ \bibnamefont {Spaldin}}, \ and\ \bibinfo
  {author} {\bibfnamefont {S.~M.}\ \bibnamefont {Selbach}},\ }\href {\doibase
  10.1557/mrc.2016.30} {\bibfield  {journal} {\bibinfo  {journal} {MRS
  Communications}\ }\textbf {\bibinfo {volume} {6}},\ \bibinfo {pages} {182}
  (\bibinfo {year} {2016})}\BibitemShut {NoStop}%
\bibitem [{\citenamefont {Rondinelli}\ \emph {et~al.}(2009)\citenamefont
  {Rondinelli}, \citenamefont {Eidelson},\ and\ \citenamefont
  {Spaldin}}]{PhysRevB.79.205119}%
  \BibitemOpen
  \bibfield  {author} {\bibinfo {author} {\bibfnamefont {J.~M.}\ \bibnamefont
  {Rondinelli}}, \bibinfo {author} {\bibfnamefont {A.~S.}\ \bibnamefont
  {Eidelson}}, \ and\ \bibinfo {author} {\bibfnamefont {N.~A.}\ \bibnamefont
  {Spaldin}},\ }\href {\doibase 10.1103/PhysRevB.79.205119} {\bibfield
  {journal} {\bibinfo  {journal} {Phys. Rev. B}\ }\textbf {\bibinfo {volume}
  {79}},\ \bibinfo {pages} {205119} (\bibinfo {year} {2009})}\BibitemShut
  {NoStop}%
\bibitem [{\citenamefont {Sakai}\ \emph {et~al.}(2011)\citenamefont {Sakai},
  \citenamefont {Fujioka}, \citenamefont {Fukuda}, \citenamefont {Okuyama},
  \citenamefont {Hashizume}, \citenamefont {Kagawa}, \citenamefont {Nakao},
  \citenamefont {Murakami}, \citenamefont {Arima}, \citenamefont {Baron},
  \citenamefont {Taguchi},\ and\ \citenamefont
  {Tokura}}]{PhysRevLett.107.137601}%
  \BibitemOpen
  \bibfield  {author} {\bibinfo {author} {\bibfnamefont {H.}~\bibnamefont
  {Sakai}}, \bibinfo {author} {\bibfnamefont {J.}~\bibnamefont {Fujioka}},
  \bibinfo {author} {\bibfnamefont {T.}~\bibnamefont {Fukuda}}, \bibinfo
  {author} {\bibfnamefont {D.}~\bibnamefont {Okuyama}}, \bibinfo {author}
  {\bibfnamefont {D.}~\bibnamefont {Hashizume}}, \bibinfo {author}
  {\bibfnamefont {F.}~\bibnamefont {Kagawa}}, \bibinfo {author} {\bibfnamefont
  {H.}~\bibnamefont {Nakao}}, \bibinfo {author} {\bibfnamefont
  {Y.}~\bibnamefont {Murakami}}, \bibinfo {author} {\bibfnamefont
  {T.}~\bibnamefont {Arima}}, \bibinfo {author} {\bibfnamefont {A.~Q.~R.}\
  \bibnamefont {Baron}}, \bibinfo {author} {\bibfnamefont {Y.}~\bibnamefont
  {Taguchi}}, \ and\ \bibinfo {author} {\bibfnamefont {Y.}~\bibnamefont
  {Tokura}},\ }\href {\doibase 10.1103/PhysRevLett.107.137601} {\bibfield
  {journal} {\bibinfo  {journal} {Phys. Rev. Lett.}\ }\textbf {\bibinfo
  {volume} {107}},\ \bibinfo {pages} {137601} (\bibinfo {year}
  {2011})}\BibitemShut {NoStop}%
\bibitem [{\citenamefont {Chen}\ and\ \citenamefont
  {Millis}(2016{\natexlab{a}})}]{PhysRevB.94.165106}%
  \BibitemOpen
  \bibfield  {author} {\bibinfo {author} {\bibfnamefont {H.}~\bibnamefont
  {Chen}}\ and\ \bibinfo {author} {\bibfnamefont {A.~J.}\ \bibnamefont
  {Millis}},\ }\href {\doibase 10.1103/PhysRevB.94.165106} {\bibfield
  {journal} {\bibinfo  {journal} {Phys. Rev. B}\ }\textbf {\bibinfo {volume}
  {94}},\ \bibinfo {pages} {165106} (\bibinfo {year}
  {2016}{\natexlab{a}})}\BibitemShut {NoStop}%
\bibitem [{\citenamefont {Gr\"unebohm}\ \emph {et~al.}(2015)\citenamefont
  {Gr\"unebohm}, \citenamefont {Marathe},\ and\ \citenamefont
  {Ederer}}]{doi:10.1063/1.4930306}%
  \BibitemOpen
  \bibfield  {author} {\bibinfo {author} {\bibfnamefont {A.}~\bibnamefont
  {Gr\"unebohm}}, \bibinfo {author} {\bibfnamefont {M.}~\bibnamefont
  {Marathe}}, \ and\ \bibinfo {author} {\bibfnamefont {C.}~\bibnamefont
  {Ederer}},\ }\href {\doibase 10.1063/1.4930306} {\bibfield  {journal}
  {\bibinfo  {journal} {Applied Physics Letters}\ }\textbf {\bibinfo {volume}
  {107}},\ \bibinfo {pages} {102901} (\bibinfo {year} {2015})}\BibitemShut
  {NoStop}%
\bibitem [{\citenamefont {Lee}\ and\ \citenamefont
  {Rabe}(2011)}]{PhysRevB.84.104440}%
  \BibitemOpen
  \bibfield  {author} {\bibinfo {author} {\bibfnamefont {J.~H.}\ \bibnamefont
  {Lee}}\ and\ \bibinfo {author} {\bibfnamefont {K.~M.}\ \bibnamefont {Rabe}},\
  }\href {\doibase 10.1103/PhysRevB.84.104440} {\bibfield  {journal} {\bibinfo
  {journal} {Phys. Rev. B}\ }\textbf {\bibinfo {volume} {84}},\ \bibinfo
  {pages} {104440} (\bibinfo {year} {2011})}\BibitemShut {NoStop}%
\bibitem [{\citenamefont {Hong}\ \emph {et~al.}(2012)\citenamefont {Hong},
  \citenamefont {Stroppa}, \citenamefont {\'I\~niguez}, \citenamefont
  {Picozzi},\ and\ \citenamefont {Vanderbilt}}]{PhysRevB.85.054417}%
  \BibitemOpen
  \bibfield  {author} {\bibinfo {author} {\bibfnamefont {J.}~\bibnamefont
  {Hong}}, \bibinfo {author} {\bibfnamefont {A.}~\bibnamefont {Stroppa}},
  \bibinfo {author} {\bibfnamefont {J.}~\bibnamefont {\'I\~niguez}}, \bibinfo
  {author} {\bibfnamefont {S.}~\bibnamefont {Picozzi}}, \ and\ \bibinfo
  {author} {\bibfnamefont {D.}~\bibnamefont {Vanderbilt}},\ }\href {\doibase
  10.1103/PhysRevB.85.054417} {\bibfield  {journal} {\bibinfo  {journal} {Phys.
  Rev. B}\ }\textbf {\bibinfo {volume} {85}},\ \bibinfo {pages} {054417}
  (\bibinfo {year} {2012})}\BibitemShut {NoStop}%
\bibitem [{\citenamefont {King-Smith}\ and\ \citenamefont
  {Vanderbilt}(1994)}]{PhysRevB.49.5828}%
  \BibitemOpen
  \bibfield  {author} {\bibinfo {author} {\bibfnamefont {R.~D.}\ \bibnamefont
  {King-Smith}}\ and\ \bibinfo {author} {\bibfnamefont {D.}~\bibnamefont
  {Vanderbilt}},\ }\href {\doibase 10.1103/PhysRevB.49.5828} {\bibfield
  {journal} {\bibinfo  {journal} {Phys. Rev. B}\ }\textbf {\bibinfo {volume}
  {49}},\ \bibinfo {pages} {5828} (\bibinfo {year} {1994})}\BibitemShut
  {NoStop}%
\bibitem [{\citenamefont {Zhong}\ \emph {et~al.}(1994)\citenamefont {Zhong},
  \citenamefont {Vanderbilt},\ and\ \citenamefont
  {Rabe}}]{PhysRevLett.73.1861}%
  \BibitemOpen
  \bibfield  {author} {\bibinfo {author} {\bibfnamefont {W.}~\bibnamefont
  {Zhong}}, \bibinfo {author} {\bibfnamefont {D.}~\bibnamefont {Vanderbilt}}, \
  and\ \bibinfo {author} {\bibfnamefont {K.~M.}\ \bibnamefont {Rabe}},\ }\href
  {\doibase 10.1103/PhysRevLett.73.1861} {\bibfield  {journal} {\bibinfo
  {journal} {Phys. Rev. Lett.}\ }\textbf {\bibinfo {volume} {73}},\ \bibinfo
  {pages} {1861} (\bibinfo {year} {1994})}\BibitemShut {NoStop}%
\bibitem [{\citenamefont {Zhong}\ \emph {et~al.}(1995)\citenamefont {Zhong},
  \citenamefont {Vanderbilt},\ and\ \citenamefont {Rabe}}]{PhysRevB.52.6301}%
  \BibitemOpen
  \bibfield  {author} {\bibinfo {author} {\bibfnamefont {W.}~\bibnamefont
  {Zhong}}, \bibinfo {author} {\bibfnamefont {D.}~\bibnamefont {Vanderbilt}}, \
  and\ \bibinfo {author} {\bibfnamefont {K.~M.}\ \bibnamefont {Rabe}},\ }\href
  {\doibase 10.1103/PhysRevB.52.6301} {\bibfield  {journal} {\bibinfo
  {journal} {Phys. Rev. B}\ }\textbf {\bibinfo {volume} {52}},\ \bibinfo
  {pages} {6301} (\bibinfo {year} {1995})}\BibitemShut {NoStop}%
\bibitem [{\citenamefont {Kresse}\ and\ \citenamefont
  {Furthm\"uller}(1996)}]{KRESSE199615}%
  \BibitemOpen
  \bibfield  {author} {\bibinfo {author} {\bibfnamefont {G.}~\bibnamefont
  {Kresse}}\ and\ \bibinfo {author} {\bibfnamefont {J.}~\bibnamefont
  {Furthm\"uller}},\ }\href {\doibase
  https://doi.org/10.1016/0927-0256(96)00008-0} {\bibfield  {journal} {\bibinfo
   {journal} {Computational Materials Science}\ }\textbf {\bibinfo {volume}
  {6}},\ \bibinfo {pages} {15 } (\bibinfo {year} {1996})}\BibitemShut {NoStop}%
\bibitem [{\citenamefont {Kresse}\ and\ \citenamefont
  {Hafner}(1994)}]{PhysRevB.49.14251}%
  \BibitemOpen
  \bibfield  {author} {\bibinfo {author} {\bibfnamefont {G.}~\bibnamefont
  {Kresse}}\ and\ \bibinfo {author} {\bibfnamefont {J.}~\bibnamefont
  {Hafner}},\ }\href {\doibase 10.1103/PhysRevB.49.14251} {\bibfield  {journal}
  {\bibinfo  {journal} {Phys. Rev. B}\ }\textbf {\bibinfo {volume} {49}},\
  \bibinfo {pages} {14251} (\bibinfo {year} {1994})}\BibitemShut {NoStop}%
\bibitem [{\citenamefont {Kresse}\ and\ \citenamefont
  {Hafner}(1993)}]{PhysRevB.47.558}%
  \BibitemOpen
  \bibfield  {author} {\bibinfo {author} {\bibfnamefont {G.}~\bibnamefont
  {Kresse}}\ and\ \bibinfo {author} {\bibfnamefont {J.}~\bibnamefont
  {Hafner}},\ }\href {\doibase 10.1103/PhysRevB.47.558} {\bibfield  {journal}
  {\bibinfo  {journal} {Phys. Rev. B}\ }\textbf {\bibinfo {volume} {47}},\
  \bibinfo {pages} {558} (\bibinfo {year} {1993})}\BibitemShut {NoStop}%
\bibitem [{\citenamefont {Bl\"ochl}(1994)}]{PhysRevB.50.17953}%
  \BibitemOpen
  \bibfield  {author} {\bibinfo {author} {\bibfnamefont {P.~E.}\ \bibnamefont
  {Bl\"ochl}},\ }\href {\doibase 10.1103/PhysRevB.50.17953} {\bibfield
  {journal} {\bibinfo  {journal} {Phys. Rev. B}\ }\textbf {\bibinfo {volume}
  {50}},\ \bibinfo {pages} {17953} (\bibinfo {year} {1994})}\BibitemShut
  {NoStop}%
\bibitem [{\citenamefont {Kresse}\ and\ \citenamefont
  {Joubert}(1999)}]{PhysRevB.59.1758}%
  \BibitemOpen
  \bibfield  {author} {\bibinfo {author} {\bibfnamefont {G.}~\bibnamefont
  {Kresse}}\ and\ \bibinfo {author} {\bibfnamefont {D.}~\bibnamefont
  {Joubert}},\ }\href {\doibase 10.1103/PhysRevB.59.1758} {\bibfield  {journal}
  {\bibinfo  {journal} {Phys. Rev. B}\ }\textbf {\bibinfo {volume} {59}},\
  \bibinfo {pages} {1758} (\bibinfo {year} {1999})}\BibitemShut {NoStop}%
\bibitem [{\citenamefont {Perdew}\ \emph {et~al.}(2008)\citenamefont {Perdew},
  \citenamefont {Ruzsinszky}, \citenamefont {Csonka}, \citenamefont {Vydrov},
  \citenamefont {Scuseria}, \citenamefont {Constantin}, \citenamefont {Zhou},\
  and\ \citenamefont {Burke}}]{PhysRevLett.100.136406}%
  \BibitemOpen
  \bibfield  {author} {\bibinfo {author} {\bibfnamefont {J.~P.}\ \bibnamefont
  {Perdew}}, \bibinfo {author} {\bibfnamefont {A.}~\bibnamefont {Ruzsinszky}},
  \bibinfo {author} {\bibfnamefont {G.~I.}\ \bibnamefont {Csonka}}, \bibinfo
  {author} {\bibfnamefont {O.~A.}\ \bibnamefont {Vydrov}}, \bibinfo {author}
  {\bibfnamefont {G.~E.}\ \bibnamefont {Scuseria}}, \bibinfo {author}
  {\bibfnamefont {L.~A.}\ \bibnamefont {Constantin}}, \bibinfo {author}
  {\bibfnamefont {X.}~\bibnamefont {Zhou}}, \ and\ \bibinfo {author}
  {\bibfnamefont {K.}~\bibnamefont {Burke}},\ }\href {\doibase
  10.1103/PhysRevLett.100.136406} {\bibfield  {journal} {\bibinfo  {journal}
  {Phys. Rev. Lett.}\ }\textbf {\bibinfo {volume} {100}},\ \bibinfo {pages}
  {136406} (\bibinfo {year} {2008})}\BibitemShut {NoStop}%
\bibitem [{\citenamefont {Dudarev}\ \emph {et~al.}(1998)\citenamefont
  {Dudarev}, \citenamefont {Botton}, \citenamefont {Savrasov}, \citenamefont
  {Humphreys},\ and\ \citenamefont {Sutton}}]{PhysRevB.57.1505}%
  \BibitemOpen
  \bibfield  {author} {\bibinfo {author} {\bibfnamefont {S.~L.}\ \bibnamefont
  {Dudarev}}, \bibinfo {author} {\bibfnamefont {G.~A.}\ \bibnamefont {Botton}},
  \bibinfo {author} {\bibfnamefont {S.~Y.}\ \bibnamefont {Savrasov}}, \bibinfo
  {author} {\bibfnamefont {C.~J.}\ \bibnamefont {Humphreys}}, \ and\ \bibinfo
  {author} {\bibfnamefont {A.~P.}\ \bibnamefont {Sutton}},\ }\href {\doibase
  10.1103/PhysRevB.57.1505} {\bibfield  {journal} {\bibinfo  {journal} {Phys.
  Rev. B}\ }\textbf {\bibinfo {volume} {57}},\ \bibinfo {pages} {1505}
  (\bibinfo {year} {1998})}\BibitemShut {NoStop}%
\bibitem [{\citenamefont {Chen}\ and\ \citenamefont
  {Millis}(2016{\natexlab{b}})}]{PhysRevB.93.205110}%
  \BibitemOpen
  \bibfield  {author} {\bibinfo {author} {\bibfnamefont {H.}~\bibnamefont
  {Chen}}\ and\ \bibinfo {author} {\bibfnamefont {A.~J.}\ \bibnamefont
  {Millis}},\ }\href {\doibase 10.1103/PhysRevB.93.205110} {\bibfield
  {journal} {\bibinfo  {journal} {Phys. Rev. B}\ }\textbf {\bibinfo {volume}
  {93}},\ \bibinfo {pages} {205110} (\bibinfo {year}
  {2016}{\natexlab{b}})}\BibitemShut {NoStop}%
\bibitem [{\citenamefont {Gajdo\ifmmode~\check{s}\else \v{s}\fi{}}\ \emph
  {et~al.}(2006)\citenamefont {Gajdo\ifmmode~\check{s}\else \v{s}\fi{}},
  \citenamefont {Hummer}, \citenamefont {Kresse}, \citenamefont
  {Furthm\"uller},\ and\ \citenamefont {Bechstedt}}]{PhysRevB.73.045112}%
  \BibitemOpen
  \bibfield  {author} {\bibinfo {author} {\bibfnamefont {M.}~\bibnamefont
  {Gajdo\ifmmode~\check{s}\else \v{s}\fi{}}}, \bibinfo {author} {\bibfnamefont
  {K.}~\bibnamefont {Hummer}}, \bibinfo {author} {\bibfnamefont
  {G.}~\bibnamefont {Kresse}}, \bibinfo {author} {\bibfnamefont
  {J.}~\bibnamefont {Furthm\"uller}}, \ and\ \bibinfo {author} {\bibfnamefont
  {F.}~\bibnamefont {Bechstedt}},\ }\href {\doibase 10.1103/PhysRevB.73.045112}
  {\bibfield  {journal} {\bibinfo  {journal} {Phys. Rev. B}\ }\textbf {\bibinfo
  {volume} {73}},\ \bibinfo {pages} {045112} (\bibinfo {year}
  {2006})}\BibitemShut {NoStop}%
\bibitem [{\citenamefont {Metropolis}\ \emph {et~al.}(1953)\citenamefont
  {Metropolis}, \citenamefont {Rosenbluth}, \citenamefont {Rosenbluth},
  \citenamefont {Teller},\ and\ \citenamefont
  {Teller}}]{doi:10.1063/1.1699114}%
  \BibitemOpen
  \bibfield  {author} {\bibinfo {author} {\bibfnamefont {N.}~\bibnamefont
  {Metropolis}}, \bibinfo {author} {\bibfnamefont {A.~W.}\ \bibnamefont
  {Rosenbluth}}, \bibinfo {author} {\bibfnamefont {M.~N.}\ \bibnamefont
  {Rosenbluth}}, \bibinfo {author} {\bibfnamefont {A.~H.}\ \bibnamefont
  {Teller}}, \ and\ \bibinfo {author} {\bibfnamefont {E.}~\bibnamefont
  {Teller}},\ }\href {\doibase 10.1063/1.1699114} {\bibfield  {journal}
  {\bibinfo  {journal} {The Journal of Chemical Physics}\ }\textbf {\bibinfo
  {volume} {21}},\ \bibinfo {pages} {1087} (\bibinfo {year}
  {1953})}\BibitemShut {NoStop}%
\bibitem [{\citenamefont {Skubic}\ \emph {et~al.}(2008)\citenamefont {Skubic},
  \citenamefont {Hellsvik}, \citenamefont {Nordstr\"{o}m},\ and\ \citenamefont
  {Eriksson}}]{Skubic2008}%
  \BibitemOpen
  \bibfield  {author} {\bibinfo {author} {\bibfnamefont {B.}~\bibnamefont
  {Skubic}}, \bibinfo {author} {\bibfnamefont {J.}~\bibnamefont {Hellsvik}},
  \bibinfo {author} {\bibfnamefont {L.}~\bibnamefont {Nordstr\"{o}m}}, \ and\
  \bibinfo {author} {\bibfnamefont {O.}~\bibnamefont {Eriksson}},\ }\href
  {\doibase 10.1088/0953-8984/20/31/315203} {\bibfield  {journal} {\bibinfo
  {journal} {Journal of Physics: Condensed Matter}\ }\textbf {\bibinfo {volume}
  {20}},\ \bibinfo {pages} {315203} (\bibinfo {year} {2008})}\BibitemShut
  {NoStop}%
\bibitem [{\citenamefont {Nishimatsu}\ \emph {et~al.}(2008)\citenamefont
  {Nishimatsu}, \citenamefont {Waghmare}, \citenamefont {Kawazoe},\ and\
  \citenamefont {Vanderbilt}}]{PhysRevB.78.104104}%
  \BibitemOpen
  \bibfield  {author} {\bibinfo {author} {\bibfnamefont {T.}~\bibnamefont
  {Nishimatsu}}, \bibinfo {author} {\bibfnamefont {U.~V.}\ \bibnamefont
  {Waghmare}}, \bibinfo {author} {\bibfnamefont {Y.}~\bibnamefont {Kawazoe}}, \
  and\ \bibinfo {author} {\bibfnamefont {D.}~\bibnamefont {Vanderbilt}},\
  }\href {\doibase 10.1103/PhysRevB.78.104104} {\bibfield  {journal} {\bibinfo
  {journal} {Phys. Rev. B}\ }\textbf {\bibinfo {volume} {78}},\ \bibinfo
  {pages} {104104} (\bibinfo {year} {2008})}\BibitemShut {NoStop}%
\bibitem [{\citenamefont {Nishimatsu}\ \emph {et~al.}(2010)\citenamefont
  {Nishimatsu}, \citenamefont {Iwamoto}, \citenamefont {Kawazoe},\ and\
  \citenamefont {Waghmare}}]{PhysRevB.82.134106}%
  \BibitemOpen
  \bibfield  {author} {\bibinfo {author} {\bibfnamefont {T.}~\bibnamefont
  {Nishimatsu}}, \bibinfo {author} {\bibfnamefont {M.}~\bibnamefont {Iwamoto}},
  \bibinfo {author} {\bibfnamefont {Y.}~\bibnamefont {Kawazoe}}, \ and\
  \bibinfo {author} {\bibfnamefont {U.~V.}\ \bibnamefont {Waghmare}},\ }\href
  {\doibase 10.1103/PhysRevB.82.134106} {\bibfield  {journal} {\bibinfo
  {journal} {Phys. Rev. B}\ }\textbf {\bibinfo {volume} {82}},\ \bibinfo
  {pages} {134106} (\bibinfo {year} {2010})}\BibitemShut {NoStop}%
\bibitem [{\citenamefont {Bond}\ \emph {et~al.}(1999)\citenamefont {Bond},
  \citenamefont {Leimkuhler},\ and\ \citenamefont {Laird}}]{BOND1999114}%
  \BibitemOpen
  \bibfield  {author} {\bibinfo {author} {\bibfnamefont {S.~D.}\ \bibnamefont
  {Bond}}, \bibinfo {author} {\bibfnamefont {B.~J.}\ \bibnamefont
  {Leimkuhler}}, \ and\ \bibinfo {author} {\bibfnamefont {B.~B.}\ \bibnamefont
  {Laird}},\ }\href {\doibase https://doi.org/10.1006/jcph.1998.6171}
  {\bibfield  {journal} {\bibinfo  {journal} {Journal of Computational
  Physics}\ }\textbf {\bibinfo {volume} {151}},\ \bibinfo {pages} {114 }
  (\bibinfo {year} {1999})}\BibitemShut {NoStop}%
\bibitem [{\citenamefont {Marathe}\ and\ \citenamefont
  {Ederer}(2014)}]{doi:10.1063/1.4879840}%
  \BibitemOpen
  \bibfield  {author} {\bibinfo {author} {\bibfnamefont {M.}~\bibnamefont
  {Marathe}}\ and\ \bibinfo {author} {\bibfnamefont {C.}~\bibnamefont
  {Ederer}},\ }\href {\doibase 10.1063/1.4879840} {\bibfield  {journal}
  {\bibinfo  {journal} {Applied Physics Letters}\ }\textbf {\bibinfo {volume}
  {104}},\ \bibinfo {pages} {212902} (\bibinfo {year} {2014})}\BibitemShut
  {NoStop}%
\bibitem [{\citenamefont {Chmaissem}\ \emph {et~al.}(2001)\citenamefont
  {Chmaissem}, \citenamefont {Dabrowski}, \citenamefont {Kolesnik},
  \citenamefont {Mais}, \citenamefont {Brown}, \citenamefont {Kruk},
  \citenamefont {Prior}, \citenamefont {Pyles},\ and\ \citenamefont
  {Jorgensen}}]{PhysRevB.64.134412}%
  \BibitemOpen
  \bibfield  {author} {\bibinfo {author} {\bibfnamefont {O.}~\bibnamefont
  {Chmaissem}}, \bibinfo {author} {\bibfnamefont {B.}~\bibnamefont
  {Dabrowski}}, \bibinfo {author} {\bibfnamefont {S.}~\bibnamefont {Kolesnik}},
  \bibinfo {author} {\bibfnamefont {J.}~\bibnamefont {Mais}}, \bibinfo {author}
  {\bibfnamefont {D.~E.}\ \bibnamefont {Brown}}, \bibinfo {author}
  {\bibfnamefont {R.}~\bibnamefont {Kruk}}, \bibinfo {author} {\bibfnamefont
  {P.}~\bibnamefont {Prior}}, \bibinfo {author} {\bibfnamefont
  {B.}~\bibnamefont {Pyles}}, \ and\ \bibinfo {author} {\bibfnamefont {J.~D.}\
  \bibnamefont {Jorgensen}},\ }\href {\doibase 10.1103/PhysRevB.64.134412}
  {\bibfield  {journal} {\bibinfo  {journal} {Phys. Rev. B}\ }\textbf {\bibinfo
  {volume} {64}},\ \bibinfo {pages} {134412} (\bibinfo {year}
  {2001})}\BibitemShut {NoStop}%
\bibitem [{\citenamefont {Wollan}\ and\ \citenamefont
  {Koehler}(1955)}]{Wollan/Koehler:1955}%
  \BibitemOpen
  \bibfield  {author} {\bibinfo {author} {\bibfnamefont {E.~O.}\ \bibnamefont
  {Wollan}}\ and\ \bibinfo {author} {\bibfnamefont {W.~C.}\ \bibnamefont
  {Koehler}},\ }\href@noop {} {\bibfield  {journal} {\bibinfo  {journal}
  {Physical Review}\ }\textbf {\bibinfo {volume} {100}},\ \bibinfo {pages}
  {545} (\bibinfo {year} {1955})}\BibitemShut {NoStop}%
\bibitem [{\citenamefont {Perdew}\ \emph {et~al.}(1996)\citenamefont {Perdew},
  \citenamefont {Burke},\ and\ \citenamefont
  {Ernzerhof}}]{PhysRevLett.77.3865}%
  \BibitemOpen
  \bibfield  {author} {\bibinfo {author} {\bibfnamefont {J.~P.}\ \bibnamefont
  {Perdew}}, \bibinfo {author} {\bibfnamefont {K.}~\bibnamefont {Burke}}, \
  and\ \bibinfo {author} {\bibfnamefont {M.}~\bibnamefont {Ernzerhof}},\ }\href
  {\doibase 10.1103/PhysRevLett.77.3865} {\bibfield  {journal} {\bibinfo
  {journal} {Phys. Rev. Lett.}\ }\textbf {\bibinfo {volume} {77}},\ \bibinfo
  {pages} {3865} (\bibinfo {year} {1996})}\BibitemShut {NoStop}%
\bibitem [{\citenamefont {Xiang}\ \emph {et~al.}(2011)\citenamefont {Xiang},
  \citenamefont {Kan}, \citenamefont {Wei}, \citenamefont {Whangbo},\ and\
  \citenamefont {Gong}}]{PhysRevB.84.224429}%
  \BibitemOpen
  \bibfield  {author} {\bibinfo {author} {\bibfnamefont {H.~J.}\ \bibnamefont
  {Xiang}}, \bibinfo {author} {\bibfnamefont {E.~J.}\ \bibnamefont {Kan}},
  \bibinfo {author} {\bibfnamefont {S.-H.}\ \bibnamefont {Wei}}, \bibinfo
  {author} {\bibfnamefont {M.-H.}\ \bibnamefont {Whangbo}}, \ and\ \bibinfo
  {author} {\bibfnamefont {X.~G.}\ \bibnamefont {Gong}},\ }\href {\doibase
  10.1103/PhysRevB.84.224429} {\bibfield  {journal} {\bibinfo  {journal} {Phys.
  Rev. B}\ }\textbf {\bibinfo {volume} {84}},\ \bibinfo {pages} {224429}
  (\bibinfo {year} {2011})}\BibitemShut {NoStop}%
\bibitem [{\citenamefont {Fedorova}\ \emph {et~al.}(2015)\citenamefont
  {Fedorova}, \citenamefont {Ederer}, \citenamefont {Spaldin},\ and\
  \citenamefont {Scaramucci}}]{PhysRevB.91.165122}%
  \BibitemOpen
  \bibfield  {author} {\bibinfo {author} {\bibfnamefont {N.~S.}\ \bibnamefont
  {Fedorova}}, \bibinfo {author} {\bibfnamefont {C.}~\bibnamefont {Ederer}},
  \bibinfo {author} {\bibfnamefont {N.~A.}\ \bibnamefont {Spaldin}}, \ and\
  \bibinfo {author} {\bibfnamefont {A.}~\bibnamefont {Scaramucci}},\ }\href
  {\doibase 10.1103/PhysRevB.91.165122} {\bibfield  {journal} {\bibinfo
  {journal} {Phys. Rev. B}\ }\textbf {\bibinfo {volume} {91}},\ \bibinfo
  {pages} {165122} (\bibinfo {year} {2015})}\BibitemShut {NoStop}%
\bibitem [{sup()}]{suppl}%
  \BibitemOpen
  \href@noop {} {}\bibinfo {note} {See Supplemental Material at [URL will be
  inserted by publisher] for exchange interactions calculated with respect to
  different magnetic structures and as function of $U$, magnon spectra and
  details regarding the evaluation of the FE effective Hamiltonian parameters.
  Additionally contains
  references~[\onlinecite{Anderson196399},\onlinecite{PhysRevB.92.024419},\onlinecite{1712.03907},\onlinecite{PhysRevB.58.293},\onlinecite{PhysRevB.64.174402},\onlinecite{Dresselhaus},\onlinecite{PhysRevB.82.134106},\onlinecite{PhysRevB.52.6301},\onlinecite{PhysRevB.49.5828}].}\BibitemShut
  {Stop}%
\bibitem [{\citenamefont {Anderson}(1963)}]{Anderson196399}%
  \BibitemOpen
  \bibfield  {author} {\bibinfo {author} {\bibfnamefont {P.~W.}\ \bibnamefont
  {Anderson}},\ }in\ \href {\doibase
  http://dx.doi.org/10.1016/S0081-1947(08)60260-X} {\emph {\bibinfo {booktitle}
  {Solid State Physics}}},\ Vol.~\bibinfo {volume} {14},\ \bibinfo {editor}
  {edited by\ \bibinfo {editor} {\bibfnamefont {F.}~\bibnamefont {Seitz}}\ and\
  \bibinfo {editor} {\bibfnamefont {D.}~\bibnamefont {Turnbull}}}\ (\bibinfo
  {publisher} {Academic Press},\ \bibinfo {year} {1963})\ pp.\ \bibinfo {pages}
  {99 -- 214}\BibitemShut {NoStop}%
\bibitem [{\citenamefont {Kvashnin}\ \emph {et~al.}(2016)\citenamefont
  {Kvashnin}, \citenamefont {Cardias}, \citenamefont {Szilva}, \citenamefont
  {Di~Marco}, \citenamefont {Katsnelson}, \citenamefont {Lichtenstein},
  \citenamefont {Nordstr\"om}, \citenamefont {Klautau},\ and\ \citenamefont
  {Eriksson}}]{PhysRevLett.116.217202}%
  \BibitemOpen
  \bibfield  {author} {\bibinfo {author} {\bibfnamefont {Y.~O.}\ \bibnamefont
  {Kvashnin}}, \bibinfo {author} {\bibfnamefont {R.}~\bibnamefont {Cardias}},
  \bibinfo {author} {\bibfnamefont {A.}~\bibnamefont {Szilva}}, \bibinfo
  {author} {\bibfnamefont {I.}~\bibnamefont {Di~Marco}}, \bibinfo {author}
  {\bibfnamefont {M.~I.}\ \bibnamefont {Katsnelson}}, \bibinfo {author}
  {\bibfnamefont {A.~I.}\ \bibnamefont {Lichtenstein}}, \bibinfo {author}
  {\bibfnamefont {L.}~\bibnamefont {Nordstr\"om}}, \bibinfo {author}
  {\bibfnamefont {A.~B.}\ \bibnamefont {Klautau}}, \ and\ \bibinfo {author}
  {\bibfnamefont {O.}~\bibnamefont {Eriksson}},\ }\href {\doibase
  10.1103/PhysRevLett.116.217202} {\bibfield  {journal} {\bibinfo  {journal}
  {Phys. Rev. Lett.}\ }\textbf {\bibinfo {volume} {116}},\ \bibinfo {pages}
  {217202} (\bibinfo {year} {2016})}\BibitemShut {NoStop}%
\bibitem [{\citenamefont {Fang}\ \emph {et~al.}(2000)\citenamefont {Fang},
  \citenamefont {Solovyev},\ and\ \citenamefont
  {Terakura}}]{PhysRevLett.84.3169}%
  \BibitemOpen
  \bibfield  {author} {\bibinfo {author} {\bibfnamefont {Z.}~\bibnamefont
  {Fang}}, \bibinfo {author} {\bibfnamefont {I.~V.}\ \bibnamefont {Solovyev}},
  \ and\ \bibinfo {author} {\bibfnamefont {K.}~\bibnamefont {Terakura}},\
  }\href {\doibase 10.1103/PhysRevLett.84.3169} {\bibfield  {journal} {\bibinfo
   {journal} {Phys. Rev. Lett.}\ }\textbf {\bibinfo {volume} {84}},\ \bibinfo
  {pages} {3169} (\bibinfo {year} {2000})}\BibitemShut {NoStop}%
\bibitem [{MFT()}]{MFTnote}%
  \BibitemOpen
  \href@noop {} {}\bibinfo {note} {To accomodate for antiferromagnetism, the
  mean field theory is done assuming a $2\times 2\times 2$ supercell
  multisublattice system (which assumes that the magnetic order fits in such a
  supercell). The mean field estimate of $T_\mathrm{C}$ for a multisublattice
  system is obtained by first diagonalising the symmetric matrix
  $J_0^{AB}=\sum_{j} J_{A_1 B_j}$, with summation over different sites $B_j$
  belonging to sublattice $B$. the mean field critical temperature is then
  $T_\mathrm{C} = \frac{J_O^\mathrm{max}}{3 k_\mathrm{B}}$, where
  $J_O^\mathrm{max}$ is the maximum eigenvalue of $J_0^{AB}$ and $k_\mathrm{B}$
  is Boltzmann's constant. See e.g.
  Refs.~\onlinecite{Anderson196399,PhysRevB.70.024427}}\BibitemShut {NoStop}%
\bibitem [{\citenamefont {\ifmmode \mbox{\c{S}}\else \c{S}\fi{}a\ifmmode
  \mbox{\c{s}}\else \c{s}\fi{}\ifmmode \imath \else \i
  \fi{}o\ifmmode~\breve{g}\else \u{g}\fi{}lu}\ \emph
  {et~al.}(2004)\citenamefont {\ifmmode \mbox{\c{S}}\else \c{S}\fi{}a\ifmmode
  \mbox{\c{s}}\else \c{s}\fi{}\ifmmode \imath \else \i
  \fi{}o\ifmmode~\breve{g}\else \u{g}\fi{}lu}, \citenamefont {Sandratskii},\
  and\ \citenamefont {Bruno}}]{PhysRevB.70.024427}%
  \BibitemOpen
  \bibfield  {author} {\bibinfo {author} {\bibfnamefont {E.}~\bibnamefont
  {\ifmmode \mbox{\c{S}}\else \c{S}\fi{}a\ifmmode \mbox{\c{s}}\else
  \c{s}\fi{}\ifmmode \imath \else \i \fi{}o\ifmmode~\breve{g}\else
  \u{g}\fi{}lu}}, \bibinfo {author} {\bibfnamefont {L.~M.}\ \bibnamefont
  {Sandratskii}}, \ and\ \bibinfo {author} {\bibfnamefont {P.}~\bibnamefont
  {Bruno}},\ }\href {\doibase 10.1103/PhysRevB.70.024427} {\bibfield  {journal}
  {\bibinfo  {journal} {Phys. Rev. B}\ }\textbf {\bibinfo {volume} {70}},\
  \bibinfo {pages} {024427} (\bibinfo {year} {2004})}\BibitemShut {NoStop}%
\bibitem [{\citenamefont {Pinettes}\ and\ \citenamefont
  {Diep}(1998)}]{10.1063/1.367729}%
  \BibitemOpen
  \bibfield  {author} {\bibinfo {author} {\bibfnamefont {C.}~\bibnamefont
  {Pinettes}}\ and\ \bibinfo {author} {\bibfnamefont {H.~T.}\ \bibnamefont
  {Diep}},\ }\href {\doibase 10.1063/1.367729} {\bibfield  {journal} {\bibinfo
  {journal} {Journal of Applied Physics}\ }\textbf {\bibinfo {volume} {83}},\
  \bibinfo {pages} {6317} (\bibinfo {year} {1998})}\BibitemShut {NoStop}%
\bibitem [{\citenamefont {Maurel}\ \emph {et~al.}(2015)\citenamefont {Maurel},
  \citenamefont {Marcano}, \citenamefont {Prokscha}, \citenamefont
  {Langenberg}, \citenamefont {Blasco}, \citenamefont {Guzm\'an}, \citenamefont
  {Suter}, \citenamefont {Mag\'en}, \citenamefont {Morell\'on}, \citenamefont
  {Ibarra}, \citenamefont {Pardo},\ and\ \citenamefont
  {Algarabel}}]{PhysRevB.92.024419}%
  \BibitemOpen
  \bibfield  {author} {\bibinfo {author} {\bibfnamefont {L.}~\bibnamefont
  {Maurel}}, \bibinfo {author} {\bibfnamefont {N.}~\bibnamefont {Marcano}},
  \bibinfo {author} {\bibfnamefont {T.}~\bibnamefont {Prokscha}}, \bibinfo
  {author} {\bibfnamefont {E.}~\bibnamefont {Langenberg}}, \bibinfo {author}
  {\bibfnamefont {J.}~\bibnamefont {Blasco}}, \bibinfo {author} {\bibfnamefont
  {R.}~\bibnamefont {Guzm\'an}}, \bibinfo {author} {\bibfnamefont
  {A.}~\bibnamefont {Suter}}, \bibinfo {author} {\bibfnamefont
  {C.}~\bibnamefont {Mag\'en}}, \bibinfo {author} {\bibfnamefont
  {L.}~\bibnamefont {Morell\'on}}, \bibinfo {author} {\bibfnamefont {M.~R.}\
  \bibnamefont {Ibarra}}, \bibinfo {author} {\bibfnamefont {J.~A.}\
  \bibnamefont {Pardo}}, \ and\ \bibinfo {author} {\bibfnamefont {P.~A.}\
  \bibnamefont {Algarabel}},\ }\href {\doibase 10.1103/PhysRevB.92.024419}
  {\bibfield  {journal} {\bibinfo  {journal} {Phys. Rev. B}\ }\textbf {\bibinfo
  {volume} {92}},\ \bibinfo {pages} {024419} (\bibinfo {year}
  {2015})}\BibitemShut {NoStop}%
\bibitem [{\citenamefont {Nakao}\ \emph {et~al.}(2014)\citenamefont {Nakao},
  \citenamefont {Yamada}, \citenamefont {Sawa}, \citenamefont {Iwasa},
  \citenamefont {Okamoto}, \citenamefont {Sudayama}, \citenamefont {Yamasaki},\
  and\ \citenamefont {Murakami}}]{NAKAO201418}%
  \BibitemOpen
  \bibfield  {author} {\bibinfo {author} {\bibfnamefont {H.}~\bibnamefont
  {Nakao}}, \bibinfo {author} {\bibfnamefont {H.}~\bibnamefont {Yamada}},
  \bibinfo {author} {\bibfnamefont {A.}~\bibnamefont {Sawa}}, \bibinfo {author}
  {\bibfnamefont {K.}~\bibnamefont {Iwasa}}, \bibinfo {author} {\bibfnamefont
  {J.}~\bibnamefont {Okamoto}}, \bibinfo {author} {\bibfnamefont
  {T.}~\bibnamefont {Sudayama}}, \bibinfo {author} {\bibfnamefont
  {Y.}~\bibnamefont {Yamasaki}}, \ and\ \bibinfo {author} {\bibfnamefont
  {Y.}~\bibnamefont {Murakami}},\ }\href {\doibase
  https://doi.org/10.1016/j.ssc.2014.01.019} {\bibfield  {journal} {\bibinfo
  {journal} {Solid State Communications}\ }\textbf {\bibinfo {volume} {185}},\
  \bibinfo {pages} {18 } (\bibinfo {year} {2014})}\BibitemShut {NoStop}%
\bibitem [{\citenamefont {{ Dresselhaus}}\ \emph {et~al.}(2008)\citenamefont {{
  Dresselhaus}}, \citenamefont {{ Dresselhaus}},\ and\ \citenamefont
  {{Jorio}}}]{Dresselhaus}%
  \BibitemOpen
  \bibfield  {author} {\bibinfo {author} {\bibfnamefont {M.~S.}\ \bibnamefont
  {{ Dresselhaus}}}, \bibinfo {author} {\bibfnamefont {G.}~\bibnamefont {{
  Dresselhaus}}}, \ and\ \bibinfo {author} {\bibfnamefont {A.}~\bibnamefont
  {{Jorio}}},\ }\href@noop {} {\emph {\bibinfo {title} {Group Theory}}}\
  (\bibinfo  {publisher} {Springer},\ \bibinfo {address} {Heidelberg},\
  \bibinfo {year} {2008})\BibitemShut {NoStop}%
\bibitem [{\citenamefont {Bhattacharjee}\ \emph {et~al.}(2009)\citenamefont
  {Bhattacharjee}, \citenamefont {Bousquet},\ and\ \citenamefont
  {Ghosez}}]{Bhattacharjee/Bousquet/Ghosez:2009}%
  \BibitemOpen
  \bibfield  {author} {\bibinfo {author} {\bibfnamefont {S.}~\bibnamefont
  {Bhattacharjee}}, \bibinfo {author} {\bibfnamefont {E.}~\bibnamefont
  {Bousquet}}, \ and\ \bibinfo {author} {\bibfnamefont {P.}~\bibnamefont
  {Ghosez}},\ }\href {\doibase 10.1103/PhysRevLett.102.117602} {\bibfield
  {journal} {\bibinfo  {journal} {Phys. Rev. Lett.}\ }\textbf {\bibinfo
  {volume} {102}},\ \bibinfo {pages} {117602} (\bibinfo {year}
  {2009})}\BibitemShut {NoStop}%
\bibitem [{\citenamefont {Ederer}\ \emph {et~al.}(2011)\citenamefont {Ederer},
  \citenamefont {Harris},\ and\ \citenamefont
  {Kov\'a\v{c}ik}}]{Ederer/Harris/Kovacik:2011}%
  \BibitemOpen
  \bibfield  {author} {\bibinfo {author} {\bibfnamefont {C.}~\bibnamefont
  {Ederer}}, \bibinfo {author} {\bibfnamefont {T.}~\bibnamefont {Harris}}, \
  and\ \bibinfo {author} {\bibfnamefont {R.}~\bibnamefont {Kov\'a\v{c}ik}},\
  }\href {\doibase 10.1103/physrevb.83.054110} {\bibfield  {journal} {\bibinfo
  {journal} {Phys. Rev. B}\ }\textbf {\bibinfo {volume} {83}},\ \bibinfo
  {pages} {054110} (\bibinfo {year} {2011})}\BibitemShut {NoStop}%
\bibitem [{\citenamefont {Keshavarz}\ \emph {et~al.}(2018)\citenamefont
  {Keshavarz}, \citenamefont {Sch\"ott}, \citenamefont {Millis},\ and\
  \citenamefont {Kvashnin}}]{1712.03907}%
  \BibitemOpen
  \bibfield  {author} {\bibinfo {author} {\bibfnamefont {S.}~\bibnamefont
  {Keshavarz}}, \bibinfo {author} {\bibfnamefont {J.}~\bibnamefont {Sch\"ott}},
  \bibinfo {author} {\bibfnamefont {A.~J.}\ \bibnamefont {Millis}}, \ and\
  \bibinfo {author} {\bibfnamefont {Y.~O.}\ \bibnamefont {Kvashnin}},\ }\href
  {\doibase 10.1103/PhysRevB.97.184404} {\bibfield  {journal} {\bibinfo
  {journal} {Phys. Rev. B}\ }\textbf {\bibinfo {volume} {97}},\ \bibinfo
  {pages} {184404} (\bibinfo {year} {2018})}\BibitemShut {NoStop}%
\bibitem [{\citenamefont {Halilov}\ \emph {et~al.}(1998)\citenamefont
  {Halilov}, \citenamefont {Eschrig}, \citenamefont {Perlov},\ and\
  \citenamefont {Oppeneer}}]{PhysRevB.58.293}%
  \BibitemOpen
  \bibfield  {author} {\bibinfo {author} {\bibfnamefont {S.~V.}\ \bibnamefont
  {Halilov}}, \bibinfo {author} {\bibfnamefont {H.}~\bibnamefont {Eschrig}},
  \bibinfo {author} {\bibfnamefont {A.~Y.}\ \bibnamefont {Perlov}}, \ and\
  \bibinfo {author} {\bibfnamefont {P.~M.}\ \bibnamefont {Oppeneer}},\ }\href
  {\doibase 10.1103/PhysRevB.58.293} {\bibfield  {journal} {\bibinfo  {journal}
  {Phys. Rev. B}\ }\textbf {\bibinfo {volume} {58}},\ \bibinfo {pages} {293}
  (\bibinfo {year} {1998})}\BibitemShut {NoStop}%
\bibitem [{\citenamefont {Pajda}\ \emph {et~al.}(2001)\citenamefont {Pajda},
  \citenamefont {Kudrnovsk\'y}, \citenamefont {Turek}, \citenamefont {Drchal},\
  and\ \citenamefont {Bruno}}]{PhysRevB.64.174402}%
  \BibitemOpen
  \bibfield  {author} {\bibinfo {author} {\bibfnamefont {M.}~\bibnamefont
  {Pajda}}, \bibinfo {author} {\bibfnamefont {J.}~\bibnamefont {Kudrnovsk\'y}},
  \bibinfo {author} {\bibfnamefont {I.}~\bibnamefont {Turek}}, \bibinfo
  {author} {\bibfnamefont {V.}~\bibnamefont {Drchal}}, \ and\ \bibinfo {author}
  {\bibfnamefont {P.}~\bibnamefont {Bruno}},\ }\href {\doibase
  10.1103/PhysRevB.64.174402} {\bibfield  {journal} {\bibinfo  {journal} {Phys.
  Rev. B}\ }\textbf {\bibinfo {volume} {64}},\ \bibinfo {pages} {174402}
  (\bibinfo {year} {2001})}\BibitemShut {NoStop}%
\end{thebibliography}%
\bibliographystyle{apsrev4-1}

\end{document}